\documentclass[useAMS,usenatbib,usegraphicx]{mn2e}
\usepackage{pdflscape}
\usepackage{multirow}
\usepackage{natbib}
\usepackage{aas_macros}
\usepackage{amsmath}

\newcommand{\gasp}{\texttt{GASP2D}}
\newcommand{\sg}{S$^4$G}
\newcommand{\STECKMAP}{\texttt{STECMAP}}

\title[TIMER double-barred galaxies]{Clocking the assembly of double-barred galaxies with the MUSE TIMER project}
\author[A. de Lorenzo-C\'aceres et al.]{
Adriana de Lorenzo-C\'aceres$^{1,2}$\thanks{E-mail:
adrianadelorenzocaceres@gmail.com}, 
Patricia S\'anchez-Bl\'azquez$^{3}$, 
Jairo M\'endez-Abreu$^{1,2}$, 
\newauthor Dimitri A. Gadotti$^{4}$, 
Jes\'us Falc\'on-Barroso$^{1,2}$,
Inma Mart\'inez-Valpuesta$^{1,2}$,
\newauthor Paula Coelho$^{5}$,
Francesca Fragkoudi$^{6}$, 
Bernd Husemann$^{7}$,
Ryan Leaman$^{7}$, 
\newauthor Isabel P\'erez$^{8,9}$,
Miguel Querejeta$^{4,10}$,
Marja Seidel$^{11,12}$,
Glenn van de Ven$^{4,7}$\\
$^{1}$Instituto de Astrof\'isica de Canarias, Calle V\'ia L\'actea s/n, E-38205 La Laguna, Tenerife, Spain\\
$^{2}$Departamento de Astrof\'isica, Universidad de La Laguna, E-38200 La Laguna, Tenerife, Spain\\
$^{3}$Departamento de F\'isica Te\'orica, Universidad Aut\'onoma de Madrid, E-28049 Cantoblanco, Spain\\
$^{4}$European Southern Observatory, Karl-Schwarzschild-Str. 2, D-85748 Garching bei Munchen, Germany\\
$^{5}$Instituto de Astronomia, Geof\'isica e Ciencias Atmosf\'ericas, Universidade de Sao Paulo, Rua do Matao, 1226, 05508-090 Sao Paulo-SP, Brazil\\
$^{6}$Max-Planck-Institut fur Astrophysik, Karl-Schwarzschild-Str. 1, D-85748 Garching bei München, Germany\\
$^{7}$Max-Planck-Institut fur Astronomie, Konigstuhl 17, D-69117 Heidelberg, Germany\\
$^{8}$Departamento de F\'isica Te\'orica y del Cosmos, Universidad de Granada, Facultad de Ciencias (Edificio Mecenas), E-18071, Granada, Spain\\
$^{9}$Instituto Universitario Carlos I de F\'isica Te\'orica y Computacional, Universidad de Granada, E-18071 Granada, Spain\\
$^{10}$Observatorio Astron{\'o}mico Nacional (IGN), C/ Alfonso XII 3, E-28014 Madrid, Spain\\
$^{11}$Caltech-IPAC, Spitzer Science Center, 1200 E. California Blvd., Pasadena, CA 91125, USA\\
$^{12}$The Observatories of the Carnegie Institution for Science, 813 Santa Barbara St., Pasadena, CA 91101, USA
}
\begin{document}
 
\date{Accepted ***. Received ***; in original form ***}

\pagerange{\pageref{firstpage}--\pageref{lastpage}} \pubyear{2018}

\maketitle

\label{firstpage}

\begin{abstract}
The formation of two stellar bars within a galaxy has proved challenging for numerical studies.
It is yet not clear whether the inner bar is born via a star formation process promoted by gas inflow
along the outer bar, or whether it is dynamically assembled from instabilities in a small-scale stellar disc. Observational constraints
to these scenarios are scarce. We present a thorough study of the stellar content of 
two double-barred galaxies observed by the MUSE TIMER project, NGC\,1291 and NGC\,5850,
combined with a two-dimensional multi-component photometric decomposition performed on the 3.6\,$\mu$m images from \sg.
Our analysis confirms the presence of $\sigma$-hollows appearing in the stellar velocity dispersion distribution
at the ends of the inner bars. Both galaxies host inner discs matching in size with 
the inner bars, suggestive of a dynamical formation for the inner bars from small-scale discs.
The analysis of the star formation histories for the structural components shaping the galaxies
provides constraints on the epoch of dynamical assembly of the inner bars, which took place $>$6.5\,Gyr ago
for NGC\,1291 and $>$4.5\,Gyr ago for NGC\,5850.
This implies that inner bars are long-lived structures.

\end{abstract}


\section{Introduction}\label{sec:intro}
Double-barred galaxies are disc galaxies which host two embedded, non-axisymmetric, 
bar-shaped stellar structures. The \emph{outer bar}
is the regular stellar bar found in 60\%-70\% of all disc galaxies
\citep[][among others]{MarinovaandJogee2007,Aguerrietal2009,MendezAbreuetal2012,MendezAbreuetal2017},
which appears alone in the majority of the galaxies.
The smaller \emph{inner bar} is sometimes referred to as a nuclear bar or secondary
bar, although we prefer avoiding the use of these terms as they hold implications for either its 
size or formation process which have not been proven yet, as explained below. The inner
bar appears randomly oriented with respect to the outer bar due to their different pattern speeds,
as found in both simulations and observations \citep{FriedliandMartinet93,Corsinietal2003}.

Current estimates indicate that 30\% of all early-type barred galaxies are actually
double-barred \citep[][]{ErwinandSparke2002, Laineetal2002, Erwin2004}. 
This fraction may even be considered a
lower limit: since inner-bar detection has traditionally relied on direct \citep[][]{Butaetal2015}
or indirect \citep[ellipse fitting, unsharp masking; ][]{Erwin2004} photometric
identification, it is hampered by the overlapping of many structural components
(bulges, inner discs, underlying outer bar and disc) in the central galaxy regions
where the inner bar is embedded. Indeed, this fact together with the presence of dust
has prevented the photometric search for double-barred systems in the latest galaxy types
\citep[beyond Sb; ][]{Erwin2005, Erwin2011}.

Outer bars are able to promote
secular evolution in a significant way, thanks to their ability of transporting material (gas and 
stars) inwards and outwards
\citep[e.g., ][]{MunozTunonetal2004,Shethetal2005}.
Inner bars have been proposed to contribute to gas inflow as well, even reaching the
regions of influence of the central black holes and igniting nuclear
activity \citep[AGN; ][]{Shlosmanetal89, Shlosmanetal90}. Although signatures of gas flow
from the outer to the inner bars have actually been detected in double-barred galaxies
\citep{deLorenzoCaceresetal2012, deLorenzoCaceresetal2013}, no clear evidence
of associated star-forming regions or newly-formed structures due to inner bars have been found yet.
The long-lived nature of inner bars is also under debate.

The two most extensive catalogs of double-barred galaxies to date are those presented by
\citet{Erwin2004} and \citet{Butaetal2015}. \citet{Erwin2004} uses a compilation of images
from different sources, including some Hubble Space Telescope (HST) data, and identifies 50
double-barred galaxies. 
\citet{Butaetal2015} performs the morphological classification
of over 2000 galaxies from the Spitzer Survey of Stellar Structure in Galaxies
\citep[S$^4$G;][]{Shethetal2010}, finding 15 double-barred galaxies. Both studies 
have nine individuals in common. We note however that 
some inner bars are misclassified with other central non-barred components, 
such as inner rings, inner discs, inner spirals, or star-forming spots that 
overlap and resemble elongated components when observed with limited spatial resolution.
In particular, the mean point-spread function (PSF) of the S$^4$G 3.6\,$\mu$m data
is 1.66\,arcsec.

Inner bars typically extend from 0.3 up to 2.5\,kpc in length (projected semi-major axes; 
de Lorenzo-C\'aceres et al., in preparation). These numbers have been for the first time
measured with two-dimensional multi-component photometric decompositions and they reveal
that inner bars may be long structures,
far from the \emph{nuclear} scales of less than 1\,kpc in full length previously considered
\citep[]{Erwin2011}. Note moreover that inner bars can be as large as some of the smallest
single bars \citep{Erwin2005}.
Notwithstanding, these measurements still mean inner bars can be as short as 
just a couple of arcsec even at the nearby distances of several known double-barred galaxies.
Spatial resolution is therefore key when detecting and analysing inner bars and it explains
the small amount of double-barred systems detected within S$^4$G.

Observational spectroscopic studies of double-barred galaxies are scarce. 
First approaches focussed on the analysis of their stellar kinematics with long-slit
 \citep{Emsellemetal2001} and integral-field spectra \citep{Moiseevetal2004}.
This last technique soon proved the best strategy to study structurally-complex objects
such as double-barred galaxies. The suite of works by \citet{deLorenzoCaceresetal2008,
deLorenzoCaceresetal2012,deLorenzoCaceresetal2013}, based on long-slit and integral-field data 
from the EMMI@NTT and SAURON@WHT spectrographs, respectively, represents the most 
extensive analysis of not only the stellar kinematics but also the stellar populations and
ionised gas content of double-barred galaxies.

\citet{deLorenzoCaceresetal2008} found that the main kinematic signature caused by inner bars
in the line-of-sight velocity distribution (LOSVD) affects the velocity dispersion. Indeed, while
the distortion of the velocity field induced by the inner bar is very subtle \citep[a slight double-hump profile
predicted by the simulations of ][]{BureauandAthanassoula2005}, the spatial distribution
of the velocity dispersion shows the so-called \emph{$\sigma$-hollows}: 
two local minima located at the ends of the inner bars. 
\citet{deLorenzoCaceresetal2008} proposed the $\sigma$-hollows appear due
to a contrast effect between the high velocity dispersion of the dynamically hotter central bulge 
and the cooler inner bar. At the bar ends, where this structure starts dominating the
galaxy light over the bright bulge, the lower $\sigma$ values become evident. More recently,
numerical simulations by \citet{Duetal2017a}
found that $\sigma$-hollows may be a physical property of all, even single, bars,
particularly related to the vertical component of the velocity dispersion of the disc where
the bar is developed. The amplitude of the hollows
is stronger for shorter bars, which explains why they have been observed only in inner bars within double-barred
systems.
Why the vertical component of $\sigma$ would be affected this way is still unknown.
We note here that $\sigma$-hollows have been observed in all the double-barred galaxies
whose stellar kinematics has been studied to date \citep[see e.g.][]{deLorenzoCaceresetal2012,
deLorenzoCaceresetal2013,Duetal2016}.

The study of the formation and evolution of double-barred galaxies with numerical simulations
has proved particularly challenging. How single bars
are created from instabilities in dynamically cold discs is a relatively well understood process
\citep[e.g.][]{Combesetal90,DebattistaandSellwood2000,Athanassoula2003}. However,
obtaining two coexistent bars in a simulated disc galaxy is far more complicated. Most of the efforts
in this direction can be classified within two main scenarios. 

The first case (hereafter \emph{scenario\,1})
includes the works by \citet[][]{FriedliandMartinet93,Helleretal2001,
ShlosmanandHeller2002, EnglmaierandShlosman2004}, among others, where
a gaseous and eventually a stellar inner bar is directly formed after gas inflow
through the outer bar. This gas is trapped in the outer-bar $x_2$ orbits
and therefore the prior formation of the large-scale bar is a requirement.
Since they are born from star formation triggered by 
gas collapse, inner bars formed this way
are expected to be younger than outer bars (with 
the magnitude of the differences depending on the timescale of the whole formation process).
Such prediction has actually been measured in observations, particularly in the spectroscopic analysis of
\citet{deLorenzoCaceresetal2013}.
The simulations corresponding to this scenario find that inner bars are transient
structures which dissolve very quickly (in few hundreds of Myr) since they do not have the dynamical 
support of, e.g., a cooler structure.
However, the large fraction of inner bars found in nature and the fact that inner bars have been observed
up to a redshift $z\sim$0.15 \citep{Liskeretal2006b}
suggest they live for long periods or, if transient, that they are 
successively destroyed and reformed in short timescales.

The second path (\emph{scenario\,2}) for double-barred formation poses that inner bars form 
dynamically from inner discs, 
in the same way as large-scale bars do, but at a smaller spatial scale.
This represents a collisionless way of forming double-barred systems, as the 
inner bar formation happens dynamically out of stars and does not involve star formation.
After first attempts by \citet{FriedliandMartinet93}, \citet{RautiainenandSalo2000}, and
\citet{ Rautiainenetal2002},
the works by \citet{DebattistaandShen2007}, \citet{ShenandDebattista2009}, and more recently \citet{Duetal2015}
successfully managed to create stable inner bars without invoking the presence of gas.
They used pure N-body simulations in which
they placed a small disc-like structure, which can be named a
disc-like bulge or inner disc, within a disc galaxy. Although the small-scale bar developed
faster from the inner disc than the large-scale bar from the main disc, the authors 
argued that small-scale discs are usually formed through secular evolution promoted by the main bar.
Outer bars are most likely formed first and, although this is not a requirement in this scenario,
it is in full agreement with the recent formation process proposed by \citet{Wozniak2015}. 
This author develops an inner bar via a stellar inner disc which is formed 
through gas inflow along the previously-formed outer bar.

Within this second scenario, there are no strict constraints on the relative ages between
inner and outer bars. They may both be coeval, as in the case of the double-barred
galaxy NGC\,357 \citep{deLorenzoCaceresetal2012}.
However, younger and more metal-rich inner bars,
as those observed by \citet{deLorenzoCaceresetal2013} and mentioned before, are also in agreement,
depending on the amount of gas present and the timescales for star formation.

One of the main differences between \emph{scenario 2} with respect to \emph{scenario 1} is that 
inner bars simulated this way can be long-lived. For example,  \citet{Wozniak2015}
forms inner bars that live for up to 7\,Gyr. 
Disentangling the nature of inner bars is very important since long-lived inner bars can be considered
serious candidates for explaining the long-debated problem of black-hole feeding.

Nowadays it is clear that spectroscopic observations and a careful analysis of the
structures inhabiting the central regions of double-barred galaxies are needed to recover their 
formation histories and finally constrain the possible formation scenarios. Our first efforts 
in \citet{deLorenzoCaceresetal2012, deLorenzoCaceresetal2013} were pioneer in this field.
Following these steps, in this paper we present the analysis of the
stellar kinematics and stellar populations of the TIMER \citep[][hereafter Paper I]{Gadottietal2019}
double-barred galaxies, which allows 
to study their properties with unprecedented detail
thanks to the high spatial resolution and large field-of-view (FoV) provided by MUSE.
We combine an analysis of the star formation histories (SFH) with the structural information provided by
two-dimensional multi-component photometric decompositions of the sample galaxies. 
This combination allows us to constrain the epoch of assembly of the different structural components of double-barred
galaxies for the first time.
While we focus here on the stellar content, the properties of the ionised gas content and gas kinematics
in double-barred galaxies are
also within the goals of TIMER and will be presented elsewhere.

A summary of the design, observations, and goals of the TIMER project is provided in 
Sect.\,\ref{sec:timer}, together with a description of the TIMER double-barred galaxies.
Sections\,\ref{sec:photdec}, \ref{sec:kin}, and \ref{sec:sfh}
describe the analysis of their photometric decompositions, stellar kinematics,
and stellar populations, respectively.
The discussion of the most notable stellar kinematic features and the formation scenario
of double-barred galaxies is presented in Sect.\,\ref{sec:discussion}.
Finally, in Sect.\,\ref{sec:conclusions} we report on the main results and conclusions of this work.

\section{The TIMER dataset}\label{sec:timer}
The {\bf T}ime {\bf I}nference with {\bf M}USE in {\bf E}xtragalactic {\bf R}ings (TIMER) project 
is a survey with the VLT-MUSE integral-field spectrograph of 24 nearby barred galaxies with 
prominent central structures, such as nuclear rings, inner spiral arms, inner bars, and inner discs 
(see Paper I). One of the project's main goals is 
to study the SFH of such structures to infer the cosmic epoch of the formation 
of the bar and the dynamical settling of the main disc. The methodology was 
demonstrated with a pilot study of NGC\,4371 \citep{Gadottietal2015}.

The TIMER sample was drawn from the S$^4$G 
\citep{Shethetal2010}, which includes only galaxies at distances below $40\,\rm{Mpc}$, 
brighter than 15.5\,B-mag, and larger than 1\,arcmin. The TIMER galaxies are all barred, 
with stellar masses above $10^{10}\,{\rm M}_\odot$ and inclinations below $\sim60^\circ$.
The presence of the bar and inner structures 
was assessed from the morphological classifications of \citet{Butaetal2015}.

Most of the observations were performed during ESO Period 97 (April to September 2016) 
with a typical seeing of $0.8-0.9$\,arcsec, mean spectral resolution of $2.65\,$\AA\ 
(full-width-at-half-maximum, FWHM), 
and spectral coverage from $4750\,$\AA\ to $9350$\,\AA. MUSE covers an almost square 
1\,arcmin$\times$1\,arcmin FoV with a contiguous spatial sampling of 0.2\,arcsec$\times$0.2\,arcsec
and a spectral sampling of $1.25$\,\AA\ per pixel.

The MUSE pipeline (version 1.6) was used to reduce the dataset 
\citep[bias, flat-fielding, wavelength and flux calibrations, sky; ][]{Weilbacheretal2012}.
We refer the reader to Paper I for further details on the sample selection, 
observations, and data reduction.

\subsection{TIMER double-barred galaxies}\label{subsec:sample}
Among the full TIMER sample, 7 out of 24 galaxies 
have been previously reported to host double bars either by \citet[][7\,galaxies]{Erwin2004}
or by \citet[][4\,galaxies]{Butaetal2015}. NGC\,1291, NGC\,1433, NGC\,5728, and NGC\,5850
are the four double-barred galaxies as classified in both studies, while \citet{Erwin2004}
includes also NGC\,1097, NGC\,4303, and NGC\,4984 in his catalog. We note, however,
that some of these inner bars may have been misclassified due to the 
presence of other central structures such as inner discs, inner spirals, inner rings, and 
star-forming regions. This is actually the case for the following galaxies, for which we
have inspected recent HST images in bandfilters covered by the TIMER spectral range:
NGC\,1097 (proposal ID: 13413; PI: K. Sheth), 
NGC\,1433 and NGC\,4303 (proposal ID: 9042; PI: S. Smartt),
NGC\,4984 (proposal ID: 15133; PI: P. Erwin, private communication), and
NGC\,5728 (proposal ID: 13755; PI: J. Greene).

NGC\,1291 and NGC\,5850 are undoubtedly double-barred hosts. 
NGC\,1291 is a closeby (distance of 8.6\,Mpc) and almost face-on ($i\sim$\,11$^{\circ}$) 
galaxy, whose inner bar is perfectly observed in direct images
without the need for any further analysis (e.g. ellipse fitting or unsharp masking). 
NGC\,5850 (distance 23.1\,Mpc; $i\sim$\,39$^{\circ}$) belongs to the sample of double-barred galaxies
analysed by \citet{deLorenzoCaceresetal2008, deLorenzoCaceresetal2013}, who
confirmed the presence of the inner bar not only photometrically but also
spectroscopically through the kinematic diagnostics of the $\sigma$-hollows (see Sect.\,\ref{sec:kin}). 
\citet{Butaetal2015} classified these two galaxias as (R)SAB(l,bl,nb)0$^{+}$ and
(R$'$)SB(r,bl,nr,nb)\underline{a}b, respectively.
NGC\,1291 and NGC\,5850 are used throughout this paper
as benchmarks to study the properties of double-barred
galaxies with the superb spatial resolution provided by MUSE.

\section{Two-dimensional multi-component photometric decompositions}\label{sec:photdec}

\begin{table}
 \centering
  \caption{Best-fitting parameters for the structural components of NGC\,1291 and NGC\,5850, obtained with the 2D photometric
decompositions performed with \gasp\ on the 3.6\,$\mu$m images.}
  \label{tab:photdec}
  \begin{tabular}{llcc}
  \hline
& & NGC\,1291 & NGC\,5850 \\
\hline
\hline
\multirow{6}{*}{Bulge}     & $\mu_e$\,(mag\,arcsec$^{-2}$) & 18.2 & 19.1 \\ 
                           & R$_e$\,(arcsec) & 9.9          & 3.6 \\ 
                           & n               & 2.97         & 2.68         \\ 
                           & b/a             & 0.95         & 0.95         \\ 
                           & PA\,($^\circ$)   & 29           & 55    \\ 
                           & B/T             & 0.184        & 0.084         \\                      
\hline                                                                                                                                                                                                                                                          
\multirow{6}{*}{Inner disc}& $\mu_e$\,(mag\,arcsec$^{-2}$) & 19.3 & 20.0 \\ 
                           & R$_e$\,(arcsec) & 15.6       & 8.0 \\ 
                           & n               & 0.92       & 1.04         \\ 
                           & b/a             & 0.96       & 0.81         \\ 
                           & PA\,($^\circ$)   & 155        & 147   \\ 
                           & D/T             & 0.101      & 0.099         \\                      
\hline                                                                                                                                                                                                                                                         
\multirow{7}{*}{Inner bar} & $\mu_o$\,(mag\,arcsec$^{-2}$) & 19.0 & 18.5 \\ 
                           & a\,(arcsec)     & 29.0       & 11.3 \\ 
                           & n               & 1.00       & 1.58          \\ 
                           & b/a             & 0.35       & 0.21          \\ 
                           & PA\,($^\circ$)   & 17         & 48  \\ 
                           & c               & 2.         & 2.           \\ 
                           & Bar/T           & 0.018      & 0.017         \\ 
\hline                                                                                                                                                                                                                                                         
\multirow{7}{*}{Outer bar} & $\mu_o$\,(mag\,arcsec$^{-1}$)  & 20.6 & 20.9 \\ 
                           & a\,(arcsec)     & 131.5      & 62.1 \\ 
                           & n               & 1.51       & 1.49        \\ 
                           & b/a             & 0.43       & 0.41        \\ 
                           & PA\,($^\circ$)   & 170        & 115  \\ 
                           & c               & 1.9        & 1.8         \\ 
                           & Bar/T           & 0.087      & 0.109        \\ 
\hline                                                                                                                                                                                                                                                         
\multirow{6}{*}{Lens}      & $\mu_e$\,(mag\,arcsec$^{-2}$) & 121.5 & N/A \\ 
                           & R$_e$\,(arcsec) & 75.3  & N/A \\ 
                           & n               & 0.61          & N/A \\ 
                           & b/a             & 0.73          & N/A \\ 
                           & PA\,($^\circ$)   & 163            & N/A \\ 
                           & L/T             & 0.206         & N/A \\                      
\hline                                                                                                                                                                                                                                                         
\multirow{5}{*}{Disc}      & $\mu_o$\,(mag\,arcsec$^{-2}$) & 21.5 & 21.3 \\ 
                           & h\,(arcsec)     & 119.1         & 52.7 \\
                           & b/a             & 0.92          & 0.80          \\ 
                           & PA\,($^\circ$)  & 160            & 160   \\ 
                           & D/T             & 0.40          & 0.692         \\ 
\hline
\end{tabular}

\begin{minipage}{8cm}
Note. The bulges are parameterised with a S\'ersic function \citep{Sersic68}
where $R_{\rm e}$, $\mu_{\rm e}$, $n$, $b/a$, $PA$, and $B/T$ are the  effective (or
half-light) radius,  the surface  brightness at  $R_{\rm e}$, the
S\'ersic index describing the  curvature of the profile, the axis ratio, 
the position angle, and the contribution of the bulge to the total
galaxy light, respectively.\vspace{0.1cm}

The galaxy discs are described with a single exponential profile
where $\mu_0$, $h$, $b/a$, $PA$, and $D/T$ are the central
surface brightness, scale-length, axis ratio, position angle, and
contribution of the disc to the total galaxy light, respectively.\vspace{0.1cm}

The bars are described with a Ferrers profile \citep{Ferrers77},
where $\mu_{\rm 0}$, $a$, $b/a$, $PA$, and $Bar/T$ represent  
the central surface  brightness, length, axis ratio, and contribution of the 
bar to the total galaxy light, respectively. 
The shape parameters are $n$, that describes the shape of the surface brightness profile, 
and $c$, that controls the shape of the isophotes following generalised ellipses 
\citep[boxy when $c>$2 vs. discy when $c<$2; ][]{Athanassoulaetal90b}.\vspace{0.1cm}

The inner discs and lens are parameterised with S\'ersic profiles analagous
to those describing the bulges.
\end{minipage}
\end{table}

\begin{figure*}
 \vspace{2pt}
 \includegraphics[angle=270., width=1.\textwidth]{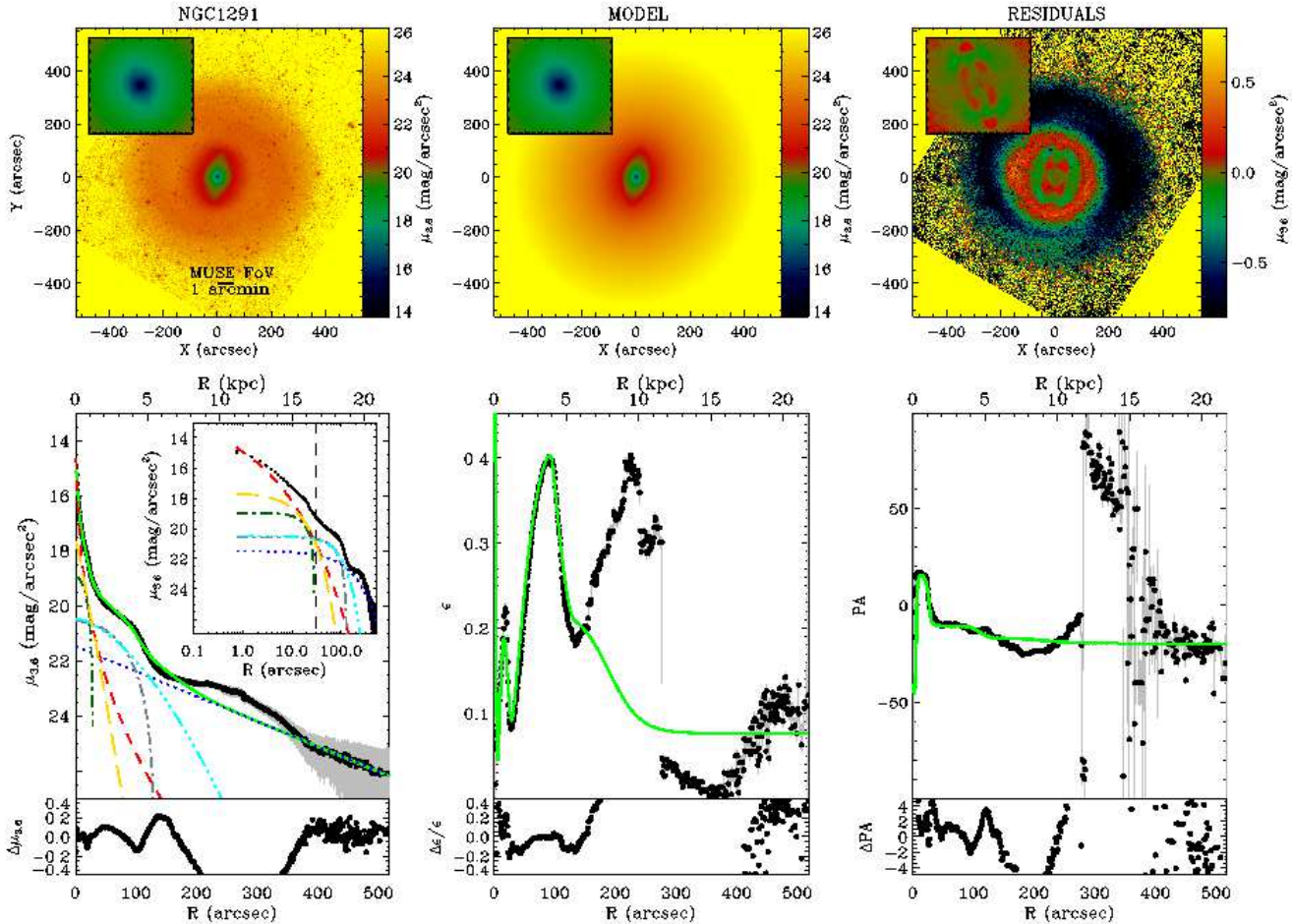}
 \caption{2D multi-component photometric decomposition performed with \gasp\ for the
double-barred galaxy NGC\,1291. The top panels show the \sg\ image, the best-fitting 2D
model, and the model$-$image residuals (from left to right). 
The insets zoom inside the corresponding MUSE FoV. 
The bottom-left panel shows the radial profile
of the image with black dots and the same measurement for the model overplotted as a green line. 
This galaxy has been fitted with 6 structural components:
bulge (red dashed line), inner bar (dark-green dash-dotted line), inner disc (yellow dashed line), 
outer bar (grey dash-dotted line), lens (cyan dash-triple-dotted line), 
and galaxy disc (blue dotted line). 
The inset shows a zoom around the central regions, with the extent 
of the MUSE FoV indicated by a vertical dashed black line.
The bottom-middle and bottom-right panels show the ellipticity and position angle profiles for the
\sg\ image (black dots) and the model (green line). All bottom panels include a lower subpanel
showing the corresponding residuals.}
 \label{fig:photdec1291}
\end{figure*}

\begin{figure*}
 \vspace{2pt}
 \includegraphics[angle=270., width=1.\textwidth]{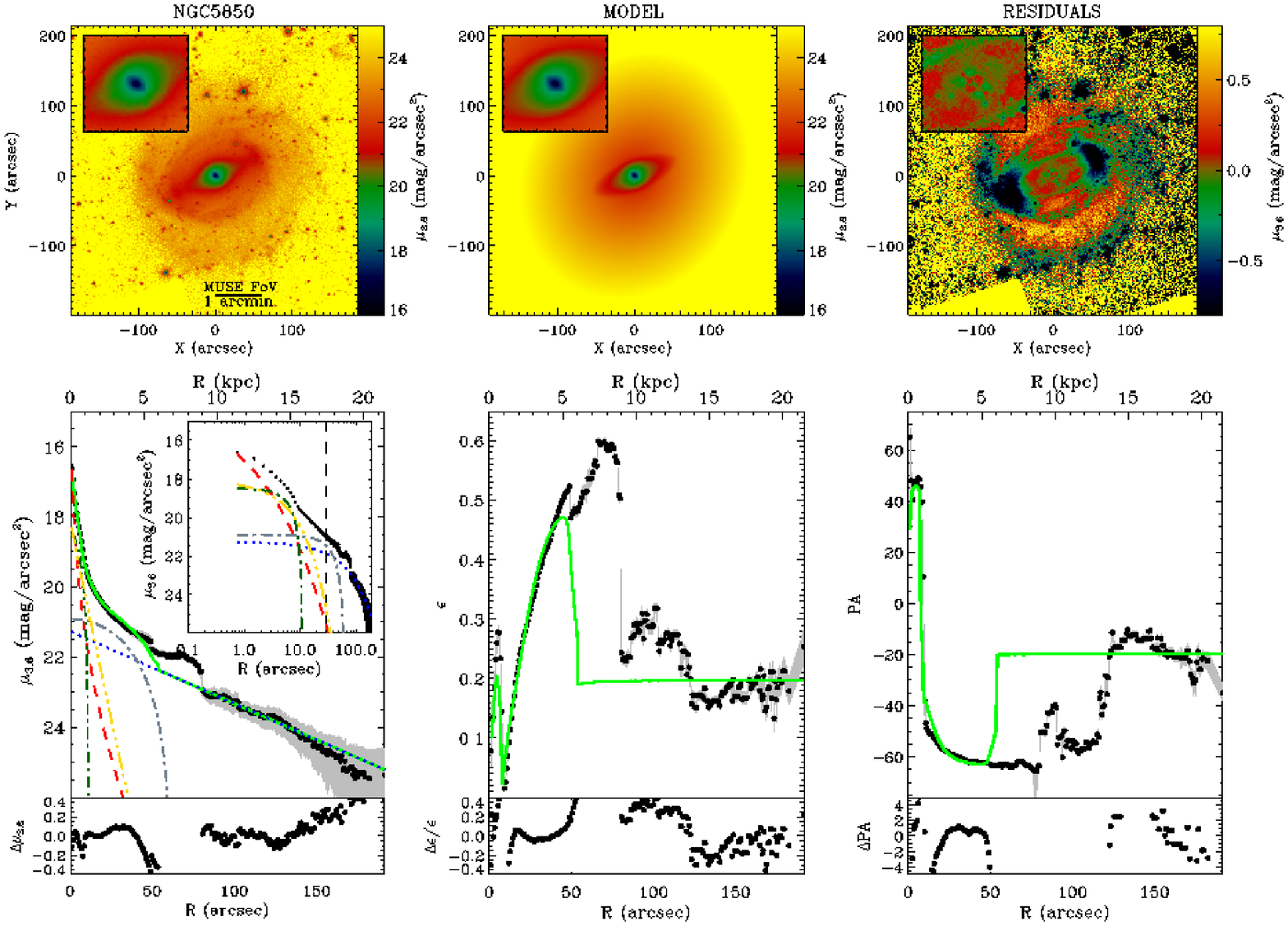}
 \caption{Same as Fig.\,\ref{fig:photdec1291} but for NGC\,5850. This galaxy hosts 5 structural
components: a bulge, 
inner bar, inner disc, outer bar, and galaxy disc. Unlike NGC\,1291, there is no outer lens.}
 \label{fig:photdec5850}
\end{figure*}

While it would be ideal to perform the photometric decompositions using the MUSE TIMER datacubes,
these spatially cover the central regions only and do not reach the galaxy disc,
whose modelling is required for a good decomposition of the central structures. 
Given the fact that all TIMER galaxies are nearby, the spatial resolution of the \sg\ images 
is good enough to derive the extent of the central structures 
including the inner bars. More importantly, \sg\ images represent a homogeneous
dataset for the whole TIMER sample and  
the 3.6\,$\mu$m bandpass has the advantage of being less affected by dust 
effects than optical ones, which is particularly important at the galaxy centres.

The multi-component photometric decomposition of the 3.6\,$\mu$m images has 
been carried out using the \gasp\ code \citep{MendezAbreuetal2008,MendezAbreuetal2014}. 
Detailed information about this analysis for the whole TIMER sample will be described in 
M\'endez-Abreu et al. (in preparation). Here we outline the basic steps  
and the specifics for the two galaxies in this paper: 
NGC\,1291 and NGC\,5850. 

\gasp\ works in two dimensions (2D) fitting the raw counts 
of the images to a combination of analytical models describing the surface-brightness 
distribution of the galaxy components. The non-linear fit is carried out minimising the 
$\chi^2$ using a Levenberg-Marquardt algorithm. The models are convolved with the image 
PSF before the $\chi^2$ is computed in each iteration. The \sg\ 3.6 $\mu$m PSF was assumed 
to be well described with a Moffat function, whose FWHM and shape parameter ($\beta$) 
were computed as the mean of several stars in each image \citep[the mathematical description of the Moffat
function can be found in, e.g., ][]{MendezAbreuetal2008}. We found FWHM values of 1.8 
and 1.9 arcsec and $\beta$ values of 2.5 and 3.0 for NGC\,1291 and NGC\,5850, respectively. 
For the fitting, each pixel was weighted according to the values provided in the weighting images supplied by 
the \sg\ team. Individual masks were created for each galaxy in order to avoid 
foreground stars or strong star-forming regions that might bias the results of the 
decomposition. This includes the possible outer and inner rings present in the images 
that were not modelled in the photometric decomposition (see below).

The final number  of components used for each galaxy  was decided in a bottom-up 
fashion  and after a careful  revision of both the  2D residuals (model$-$image)  
and the one-dimensional (1D) azimuthally-averaged radial  profiles of the surface 
brightness, ellipticity, and position angle 
(see Fig.\,\ref{fig:photdec1291} and \ref{fig:photdec5850}). Since all galaxies in  the  TIMER 
sample  host  a clear  bar,  the initial  fit  was performed using  a bulge, 
a  disc, and a  bar component described with  a S\'ersic \citep{Sersic68},     
exponential     \citep{Freeman70},    and    Ferrers \citep{Ferrers77}  profiles, respectively.  
The  inner  bars present  in  NGC\,1291  and NGC\,5850 stand out  clearly in the images 
and therefore  they were also modelled  using  an additional  Ferrers  profile. 
A complete description of the adopted functions for double-barred galaxies is detailed in \citet{deLorenzoCaceresetal2019a}.
It is worth noting that the Ferrers profiles are implemented into a reference frame of generalised ellipses, 
following the description given in \citet{Athanassoulaetal90b}.
Finally, we  included  an inner disc   component    (described    with   a    
S\'ersic    profile) in the  case of NGC\,5850, and both  an inner disc 
and outer lens  (each described with  a S\'ersic  profile) in the  case of NGC\,1291. 
In summary, we fit 6 and 5 structural components to the surface-brightness 
distributions of NGC1291 and NGC5850, respectively. 
The residuals inside the region
corresponding to the MUSE FoV represent only the 3\% and 6\% of the total flux for NGC\,1291 and NGC\,5850, respectively,
with 99.5\% of the pixels having relative errors $<$15\%. These numbers support
the robustness of our photometric decompositions to describe the structural compositions of
both galaxies within the region under study.

Our strategy 
of building up successively more complex models starting from relatively 
simple ones was successfully tested in \citet{MendezAbreuetal2017}, and it allows to 
gradually fix some parameters and reduce the highly degenerate parameter space. 
The results from our 2D photometric decompositions are shown in Table\,\ref{tab:photdec},
and in Fig.\,\ref{fig:photdec1291} and \ref{fig:photdec5850}. 
The complex morphology of the central regions of the galaxies is well captured by the 
models and resembles the visual classification given by \citet[][]{Butaetal2015}.
A notable exception are the galaxy rings and spiral arms, which were not included in the fit, but masked 
out instead. The bumps in ellipticy shown in Fig.\,\ref{fig:photdec1291} (at $\sim$9\,kpc)
and \ref{fig:photdec5850} (at $\sim$12\,kpc) correspond to these non-fitted features. 
In addition, \citet{Butaetal2015} considered the presence of a barlens 
in NGC\,5850 that we call inner disc. The interpretation of this 
component in terms of the fitting function is ambiguous, as we use a S\'ersic profile for fitting both 
barlenses and discs in the modelling. However, the MUSE TIMER velocity field clearly shows the presence of a 
fast-rotating component at the same spatial position as the inner disc (see Sect.\,\ref{sec:kin}).
The case of the inner disc in NGC\,1291 is less clear in the velocity field,
probably because it is seen face-on (see discussion in Sect.\,\ref{sec:kin}).

\section{Stellar kinematics}\label{sec:kin}

\begin{figure*}
 \vspace{2pt}
 \includegraphics[bb= 44 10 660 400, angle=0., width=.95\textwidth]{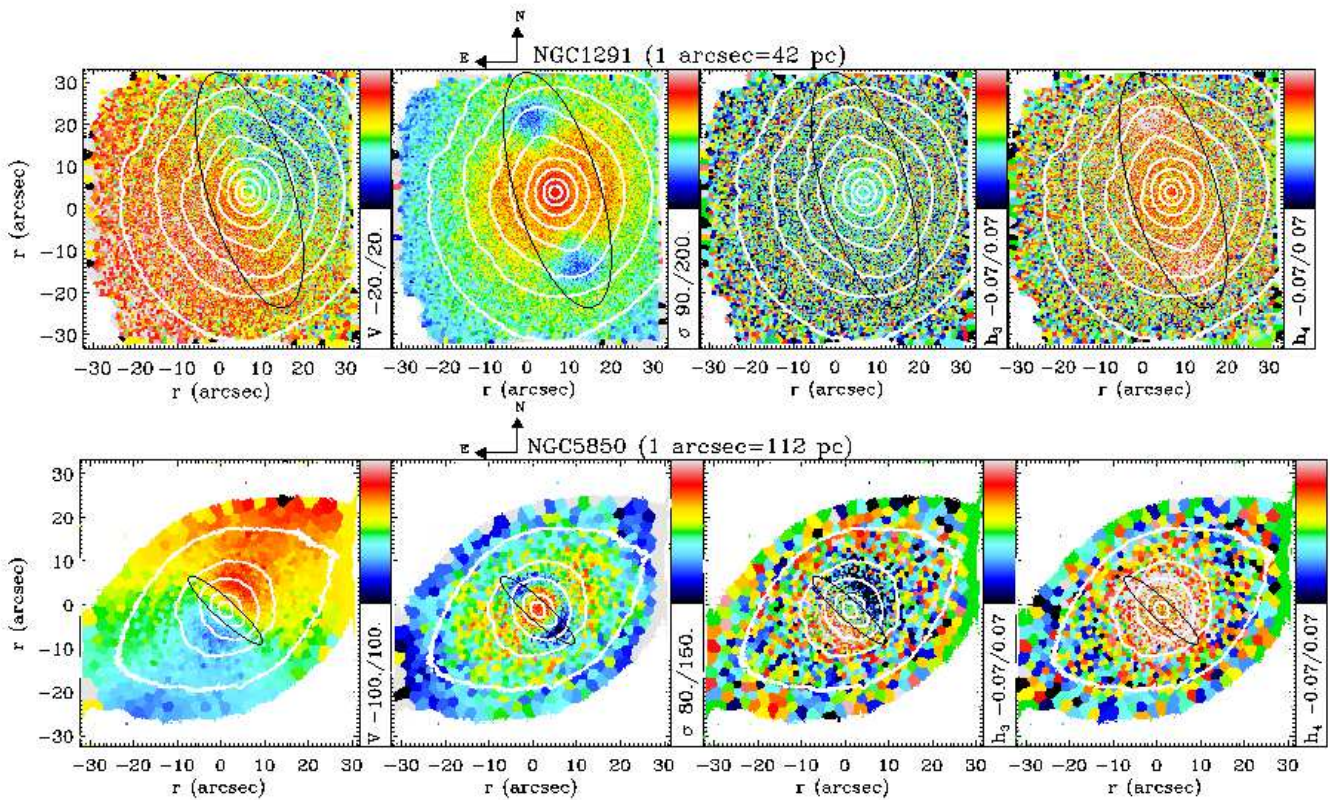}
 \caption{2D distribution of the line-of-sight stellar kinematic moments for 
NGC\,1291 (top panels) and NGC\,5850 (bottom panels). For each galaxy and from left to right, the subpanels correspond to 
the stellar velocity $V$ in km\,s$^{-1}$, velocity dispersion $\sigma$ in km\,s$^{-1}$, h$_3$, and h$_4$. 
Median errors for the four parameters are
$[$5\,km\,s$^{-1}$, 6\,km\,s$^{-1}$, 0.03, 0.03$]$ for NGC\,1291 
and $[$4\,km\,s$^{-1}$, 6\,km\,s$^{-1}$, 0.03, 0.04$]$ for NGC\,5850. Flux contours are overplotted as solid white 
lines; the inner bar (full length of the Ferrers profile) is outlined in black 
using the results from the 2D photometric decomposition. 
The presence of the inner bars is clearly noticeable in the velocity dispersion maps, where 
two $\sigma$-hollows appear at both ends of the inner bars. The inner disc in NGC\,5850 
is revealed by the velocity and h$_3$ distributions.}
 \label{fig:kin}
\end{figure*}

The 2D distributions of the stellar velocity, velocity dispersion, and higher-order
Gauss-Hermite moments h$_3$ and h$_4$ 
\citep[][]{Gerhard93,vanderMarelandFranx93} 
have been computed for all the TIMER galaxies, as
explained in Paper I and \citet{Gadottietal2015}. 
After Voronoi binning the datacubes to ensure a minimum signal-to-noise ratio SNR$>$40 per pixel
(note that the original TIMER datacubes reach SNR of 136 and 67 per pixel for NGC\,1291 and NGC\,5850, respectively),
the \texttt{pPXF} code \citep{CappellariandEmsellem2004} was used in combination with the single stellar
population (SSP) E-MILES models from \citet{Vazdekisetal2012,Vazdekisetal2016} in order to fit the
LOSVD for each spectrum corresponding to each bin. Potential emission lines
were previously masked. More details on the binning procedure, the spectral range used for
the fits, and the SSP models can be found in Sect.\,4.1 of Paper I.

Figure\,\ref{fig:kin} shows the four velocity moments for the two double-barred galaxies:
NGC\,1291 and NGC\,5850. As expected, the presence of the inner bar is barely noticeable in the
stellar velocity distribution, with only a minor distortion of the isovelocity contours
observed in NGC\,5850.
NGC\,1291 is a nearly face-on galaxy (inclination $i=$11$^{\circ}$, see Sect.\,\ref{subsec:sample}) 
and its inner bar is almost aligned with the line of nodes; in this case, no 
distortion of the isovelocity contours is expected \citep{Duetal2016}.

The most remarkable kinematic features are the 
$\sigma$-hollows, seen as blue regions at the ends of
both inner bars in the corresponding top-right subpanels of Fig.\,\ref{fig:kin}. 
The $\sigma$-hollows have amplitudes of
$\Delta\,\sigma\,\sim\,-$40\,km\,s$^{-1}$ and $\sim-$30\,km\,s$^{-1}$ for NGC\,1291 and NGC\,5850, respectively; 
these values are in full agreement
with the results found in \citet{deLorenzoCaceresetal2008} for NGC\,5850 and the rest
of their sample, for which the hollows showed amplitudes between $-$40\,km\,s$^{-1}$ and $-$10\,km\,s$^{-1}$. 

We note that $\sigma$-hollows have been observed in other  
double-barred galaxies \citep[][]{deLorenzoCaceresetal2012,
deLorenzoCaceresetal2013,Duetal2016} and they seem, so far,
an ubiquitous property of these systems. This 
result reinforces the validity of $\sigma$-hollows
as kinematic diagnostics for identifying double-barred galaxies.

Apart from the inner bar, NGC\,5850 hosts a kinematically decoupled inner disc
as seen in the velocity map and confirmed by the expected $V$-h$_3$ anti-correlation
\citep[witnessed as more intense -and opposite- red and blue regions in the 
corresponding subpanels in Fig.\,\ref{fig:kin}; e.g.][]{BureauandAthanassoula2005}. 
h$_3$ measures asymmetric deviations from the pure Gaussian behaviour of the LOSVD;
a rotating disc generates a more extended LOSVD towards low-velocity values and
this is therefore detected with the h$_3$ moment.
The kinematically-observed inner disc in NGC\,5850 corresponds to the inner disc revealed
by the photometric analysis described in Sect.\,\ref{sec:photdec}; both analyses
were performed independently and retrieved matching sizes for this inner component.
The inner disc revealed by the photometric decomposition in NGC\,1291
has no clear kinematic counterpart; this is not surprising as the almost face-on
nature of this galaxy highly hampers the detection of rotational velocity features.

The higher-order Gauss-Hermite moment h$_4$ describes symmetric deviations of the LOSVD from
a perfect Gaussian. The h$_4$ distribution for NGC\,5850 shows a ring-shaped feature
with positive values (white regions in the bottom-right subpanel of Fig.\,\ref{fig:kin}). 
These kinds of rings have been found to be a kinematic property
of inner bars seen almost face-on by means of numerical simulations \citep{Duetal2016, ShenandDebattista2009}.
Figure\,10 in \citet{Duetal2016}
shows that the presence of an inner bar-associated h$_4$ ring is noticeable in galaxies with inclinations of
$i\sim$30$^{\circ}$, similar to that of NGC\,5850 ($i=$39$^{\circ}$). 
Note that such h$_4$ feature does not correspond to an actual ringed structure 
in the galaxy: it is a pure kinematic property
which may indicate that inner bars are thinner than their surroundings.
While the presence of the 
kinematically decoupled inner disc in NGC\,5850 is already discussed in \citet{deLorenzoCaceresetal2013},
its positive-h$_4$ ring has only been unveiled thanks to the resolution the TIMER data. This demonstrates the
necessity of both high spatial resolution and SNR to study the behaviour of these
high-order moments which span short dynamical ranges.

NGC\,1291 does not show the positive-h$_4$ ring,
but two absolute (within the MUSE FoV) maxima coincident
with the position of the $\sigma$-hollows. These are seen as 
two white blobs in the h$_4$ distribution of NGC\,1291 shown in Fig.\,\ref{fig:kin}.
The h$_4$ distribution in NGC\,1291, 
suggestive of the presence of a box/peanut structure, 
is the subject of a companion TIMER paper \citep{MendezAbreuetal2019}.

\section{Stellar populations}\label{sec:sfh}

\subsection{Full-spectrum-fitting with \STECKMAP}\label{sec:steckmap}

\begin{figure*}
 \vspace{2pt}
 \includegraphics[angle=0., bb= 54 30 620 580, width=.7\textwidth]{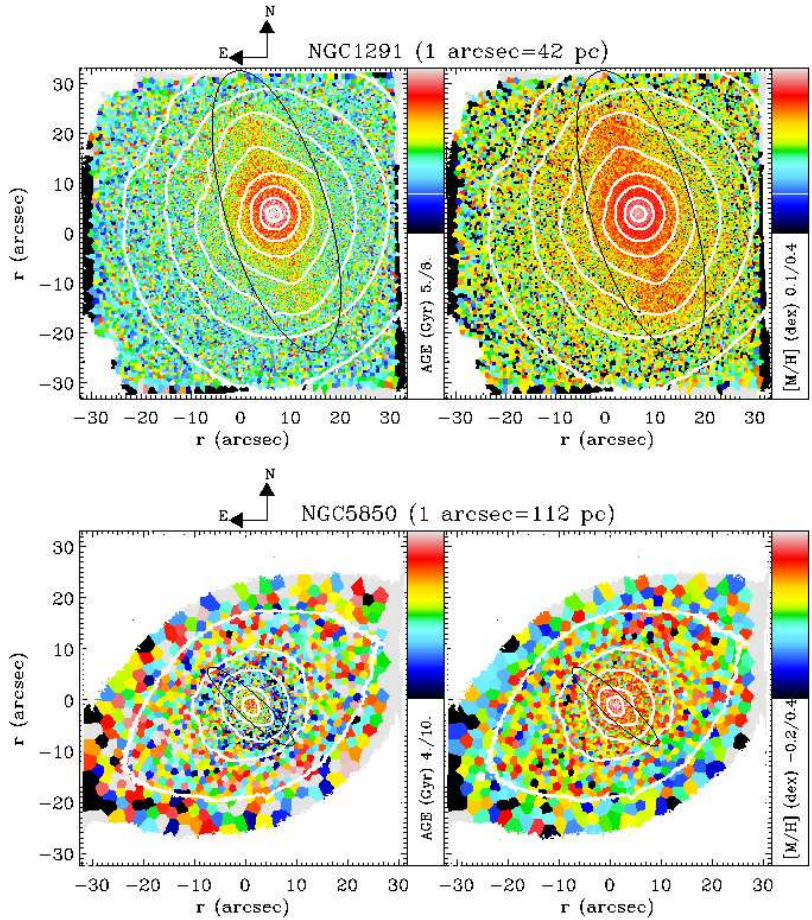}
 \caption{2D distributions of the mean luminosity-weighted ages (left subpanels) and metallicities (right subpanels)
for NGC\,1291 (on top) and NGC\,5850 (on bottom). Median error bars
are of 0.9\,Gyr and 0.04\,dex for NGC\,1291, and of 0.7\,Gyr and 0.06\,dex for NGC\,5850. Flux contours 
are overplotted as solid white lines, and the inner bar (full length of the 
Ferrers profile) is outlined in black 
by using the results from the 2D photometric decomposition. Different age and metallicity ranges
are used for each galaxy to enhance the various features. The inner bar in NGC\,1291 appears older and 
more metal-rich than the surroundings, the inner bar in NGC\,5850 being younger and more metal-rich.}
 \label{fig:meanageandmet}
\end{figure*}

The analysis of the stellar populations of the TIMER galaxies has been carried out with
the \STECKMAP\ code 
\citep[STEllar Content via Maximum A Posteriori;][]{Ocvirketal2006a}, 
following the procedure
extensively described in Sect.\,5.1 of Paper I and in S\'anchez-Bl\'azquez et al. (in preparation).
Although we refer the reader to those papers for details on the use of \STECKMAP\ within TIMER, we 
repeat here the most relevant aspects of the process. 
First, we remark that the stellar-population analysis has been performed
on the Voronoi-binned data described in Sect.\,\ref{sec:kin}, and that the gaseous contributions
have been previously removed with \texttt{GANDALF} \citep{Sarzietal2006} in order to study pure stellar spectra.

\STECKMAP\ fits every spectrum corresponding to each bin with a linear combination
of SSP models and a normalisation curve. 
Note that, although fitting the kinematics simultaneously with the stellar 
populations is among the capabilities of \STECKMAP\ \citep[in its version \texttt{STECKMAP}; ][]{Ocvirketal2006b}, we have kept the stellar kinematics
fixed to the results obtained with \texttt{pPXF}. This is done
to avoid degeneracies between the derived metallicity and the velocity dispersion 
\citep[][]{Kolevaetal2007,SanchezBlazquezetal2011}.
As templates, we have again made use of the extended version of the MILES-based
\citep{SanchezBlazquezetal2006, FalconBarrosoetal2011}
models, E-MILES 
\citep[][]{Vazdekisetal2012,Vazdekisetal2016},
which cover the full spectral range of the TIMER data. In particular, models
built with the BaSTI isochrones \citep[][and references therein]{Pietrinfernietal2013} 
and following a \citet{Kroupa2001} initial mass function are employed.
Both data and templates are previously convolved to a constant spectral resolution of 
2.8\,\AA, the maximum value of the wavelength-dependent MUSE resolution.

\STECKMAP\ allows the user to add a penalising term to the minimising function 
in order to regularise the solution. This penalisation is used to constrain
the best-fitting solution within physically meaningful limits, since retrieving the
stellar populations with full-spectrum-fitting techniques is a highly-degenerated and ill-conditioned
problem. The choice of the most suitable regularisation for the TIMER spectra
is described in Paper I and S\'anchez-Bl\'azquez et al. (in preparation).

After running \STECKMAP\ on the TIMER datacubes, we obtain
the relative contributions of populations of stars with different ages and 
metallicities within each Voronoi bin. 
Figure\,\ref{fig:meanageandmet}
shows the 2D spatial distributions of the  mean luminosity-weighted ages and metallicities 
for NGC\,1291 and NGC\,5850, computed by using
the flux from the whole MUSE spectral range in each bin as weights. 
Both measurements have been averaged in linear scale.
Note that, for the sake of enhancing all features, 
different age and metallicity ranges are shown
for each galaxy in Fig.\,\ref{fig:meanageandmet}. Mean mass-weighted values have been
obtained as well, providing analogous 2D distributions but
with older values, as expected. Light-weighted measurements are used in the following
as they enhance better the features of interest.

Results for NGC\,5850 are in good agreement with the mean luminosity-weighted ages
and metallicities derived through measurement of line-strength indices in
\citet{deLorenzoCaceresetal2013}: the inner bar  appears younger 
($\sim$5\,Gyr)
and more metal-rich than the surrounding regions corresponding
to the outer bar \citep[note that, due to the FoV, we did not probe beyond the outer bar in ][]{deLorenzoCaceresetal2013}. 
However, the higher spatial resolution
provided by MUSE with respect to the SAURON IFU spectrograph used in 
\citet{deLorenzoCaceresetal2013} reveals more details on the complex
age distribution inside and around the inner bar of NGC\,5850. 
This will be further discussed in 
Sect.\,\ref{sec:formation}.

Unlike all double-barred galaxies for which analysis of stellar populations
is available to date, which show that inner bars are younger and more metal-rich than outer bars
\citep[][]{deLorenzoCaceresetal2012, deLorenzoCaceresetal2013},
NGC\,1291 hosts an inner bar which is older than the outer bar.
Even though the dynamical range for the ages in this galaxy is very limited
(the colour bar in Fig.\,\ref{fig:meanageandmet} covers from 5 to 8\,Gyr, with a
characteristic error bar of $\sim$0.9\,Gyr),
the mean luminosity-weighted age map clearly traces the 
inner-bar shape with older ages than the surroundings. Such an effect is hardly due to errors in the measurements alone.
This surprising result will be discussed in Sect.\,\ref{sec:formation}.

The metallicity map for NGC\,1291 does show the higher values for the inner bar found for other
double-barred cases
\citep[][although, once again, note the short dynamical range of this parameter]{deLorenzoCaceresetal2013}.
For both galaxies,
the very centres within the inner bars, where the metallicity reaches the highest
values, show old ages up to 9\,Gyr. 

\subsection{Dissecting the galaxies: segmentation maps}\label{sec:segmentationmaps}

\begin{figure*}
 \vspace{2pt}
 \includegraphics[bb = 54 30 564 650,angle=0., width=.8\textwidth]{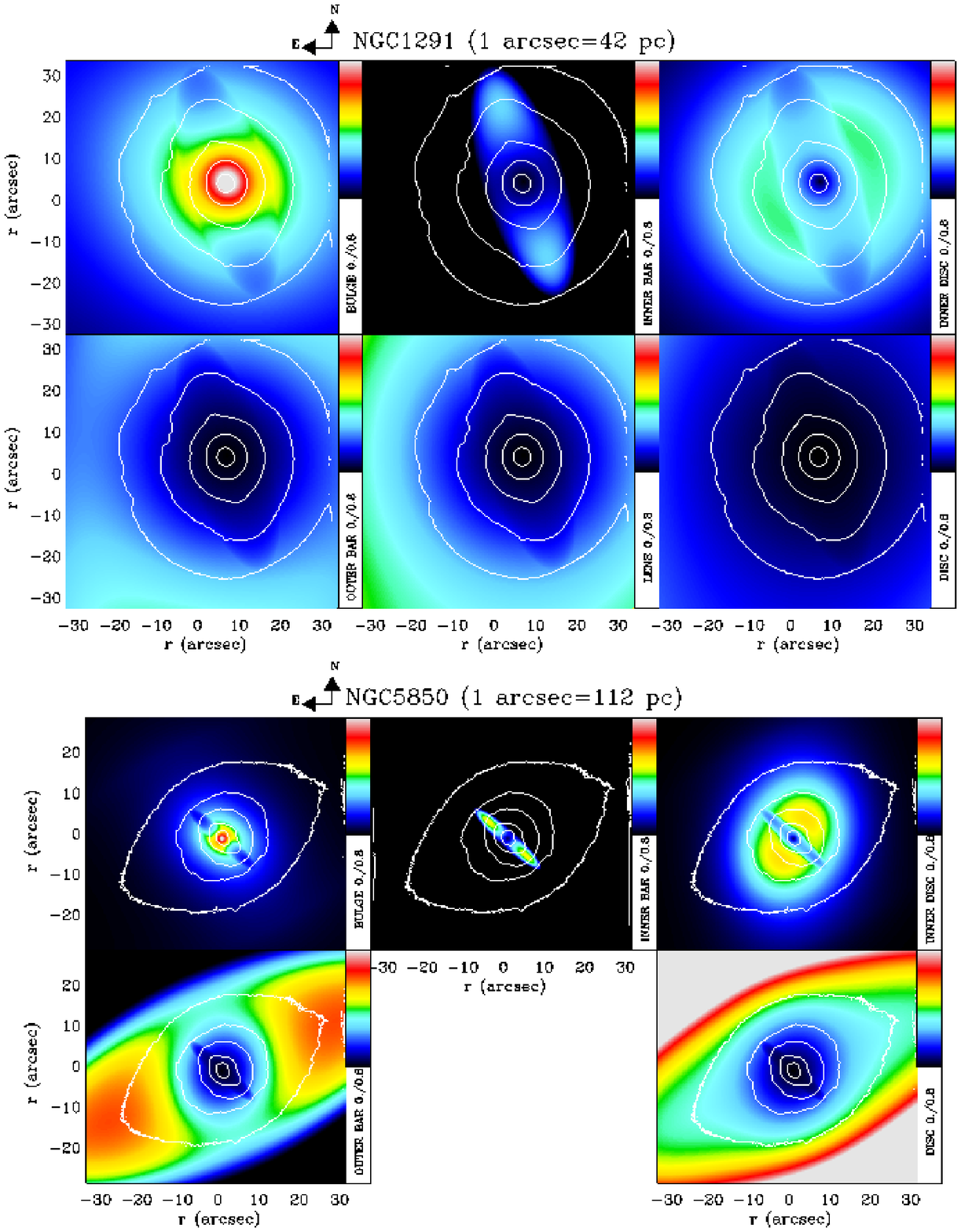}
 \caption{Spatial distribution of the relative light contribution 
from each structural component to the total surface brightness of each galaxy (NGC\,1291 in the top panels;
NGC\,5850 in the bottom panels) \emph{within each spaxel}, as modelled with the 2D 
photometric decompositions described in Sect.\,\ref{sec:photdec}: bulge
(top-left subpanel), inner bar (top-middle subpanel), inner disc (top-right subpanel),
outer bar (bottom-left subpanel), lens (bottom-middle subpanel, when applicable),
and galaxy disc (bottom-right subpanel). For the sake of clarity, all subpanels show values 
ranging from 0\% to 80\% 
of light contribution from each structure to the total galaxy light at each spaxel.
The addition of all panels would provide a flat image of value 1 (i.e., 100\%).
The images match the FoV and spatial resolution of the TIMER data for these galaxies.
The Voronoi binning is not applied at this stage.
Flux contours for the actual MUSE data are overplotted as white solid lines.}
 \label{fig:structuremaps}
\end{figure*}

\begin{table*}
 \centering
  \caption{Selection of the MUSE bins whose light is dominated by every structural component in NGC\,1291. 
Each row corresponds to one individual component. The first two columns indicate the minimum and maximum
relative contribution from that component to the total galaxy light \emph{within the FoV}. The remaining columns
indicate the criteria of light contribution from all the components 
used for creating the segmentation map shown in Fig.\,\ref{fig:sfh1291} (top-left panel). 
Two bulge-dominated regions are considered for this galaxy (see text for details). The disc
never dominates within the MUSE FoV.}
  \label{tab:structures1291}
  \begin{tabular}{lcccccccc}
  \hline
Structure  & Min \% light & Max \% light & Bulge  & Inner bar & Inner disc & Outer bar & Lens & Disc\\
\hline
\hline   
INNER BULGE & 7\% & 97\% & {\bf $\geq$60\% and $\leq$75} & $<$60\%         &  $<$60\%             &  $<$60\%             & $<$60\%          &  $<$60\%   \\
OUTER BULGE & 7\% & 97\% & {\bf $\geq$45\%} & $<$0.5\%         &  $<$45\%             &  $<$45\%             & $<$45\%          &  $<$45\%   \\
INNER BAR   & 0\% & 27\% & $<$28\%          & {\bf $\geq$25\%} & $<$28\%              & $<$25\%              & $<$25\%          & $<$25\%    \\
INNER DISC  & 2\% & 38\% & $<$35\%          & $<$35\%          & {\bf $\geq$35\%}     & $<$35\%              & $<$35\%          & $<$35\%    \\
OUTER BAR   & 0\% & 35\% & $<$25\%          & $<$25\%          & $<$25\%              & {\bf $\geq$25\%}     & $<$27\%          & $<$25\%    \\
LENS        & 0\% & 51\% & $<$30\%          & $<$30\%          & $<$30\%              & $<$30\%              & {\bf $\geq$30\%} & $<$30\%    \\
DISC        & 0\% & 24\% & -                & -                & -                    & -                    & -                & -          \\
\hline
\end{tabular}
\end{table*}

\begin{table*}
 \centering
  \caption{Same as Table\,\ref{tab:structures1291} but for NGC\,5850. 
The corresponding segmentation map is shown in Fig.\,\ref{fig:sfh5850}
(top-left panel).}
  \label{tab:structures5850}
  \begin{tabular}{lccccccc}
  \hline
Structure  & Min \% light & Max \% light & Bulge  & Inner bar & Inner disc & Outer bar & Disc\\
\hline
\hline   
BULGE       & 0\% & 91\% & {\bf $\geq$35\%} & $<$0.5\%         & $<$50\%              & $<$35\%              & $<$35\%    \\
INNER BAR   & 0\% & 48\% & $<$40\%          & {\bf $\geq$40\%} & $<$40\%              & $<$40\%              & $<$40\%    \\
INNER DISC  & 0\% & 52\% & $<$20\%          & $<$40\%          & {\bf $\geq$40\%}     & $<$40\%              & $<$40\%    \\
OUTER BAR   & 0\% & 63\% & $<$40\%          & $<$40\%          & $<$10\%              & {\bf $\geq$50\%}     & $<$40\%    \\
DISC        & 0\% & 99\% & $<$40\%          & $<$40\%          &$<$40\%               & $<$40\%              & {\bf $\geq$50\%} \\
\hline
\end{tabular}
\end{table*}

Significant information is contained in the best-fitting solutions provided by \STECKMAP\ for the stellar
population content of the TIMER galaxies.
Azimuthal profiles are not a good approach as the structural complexity
of the central regions of double-barred galaxies makes the isophotal ellipses round
due to the overlapping of many components.
As a result, the azimuthally-averaged profile will mix information from various structures. 
This is well illustrated with the case of the inner bar of NGC\,1291, for which the 
isodensity contours plotted over
the maps throughout this paper (white lines) are much rounder than the 
inner bar alone (black lines).

Since we aim at 
discussing the structural assembly history of double-barred systems with the SFH
results from \STECKMAP, our strategy consists in comparing the measurements within inner bars with
those within other components, namely bulges, outer bars, inner discs, lenses, and discs. 
For this purpose, we average the relative contributions of the different stellar populations 
over apertures where the light contribution of each structure
is mostly dominant. This approach is particularly important in structurally complex
galaxies such as those presented here.

Figure\,\ref{fig:structuremaps} shows the relative light contribution from each
structural component to every MUSE spaxel for the two galaxies under study. These maps
are built by taking the analytical description of each structure (obtained from the best fit of the 2D
photometric decompositions described in Sect.\,\ref{sec:photdec}) and projecting it onto a MUSE
FoV with the MUSE spatial resolution. The total addition
of these maps (whose minimum and maximum values are indicated in 
Tables\,\ref{tab:structures1291} and \ref{tab:structures5850}) generates a flat image with a value of 1.

We must note here that the effect of the MUSE PSF is not taken into account
in the structure maps shown in Fig.\,\ref{fig:structuremaps}:
the PSF affects the total galaxy image, but it has no physical meaning to
apply it to each 2D structure model separately. 
Dismissing the PSF has a highest impact on the recovery of the light contribution 
of the very central
structures (mainly bulge) and, for this reason, very central regions are considered
separately in the following analysis. The characteristic PSF-FWHM
for each galaxy
is computed as the median value for all the individual exposures involved in 
the final TIMER datacube and it acquires a value of 1.1\,arcsec and 1\,arcsec
for NGC\,1291 and NGC\,5850, respectively.

\begin{figure*}
 \includegraphics[bb= 10 25 479 440]{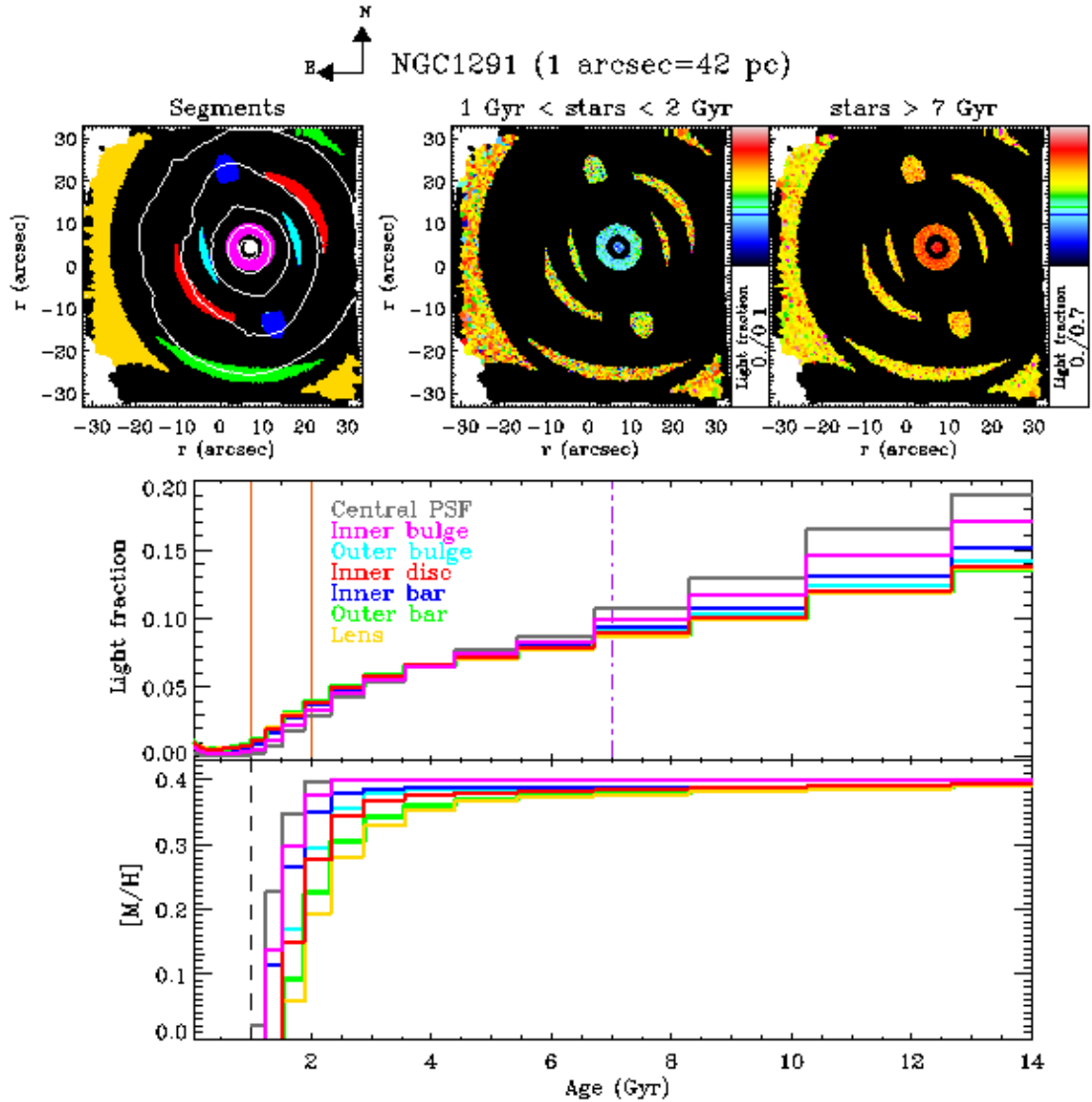}
 \caption{Analysis of the SFH for the double-barred galaxy NGC\,1291.
The top-left panel is the segmentation map defining the regions where each structural component dominates
the total galaxy light (see text for details). Flux contours for the actual MUSE data are overplotted as white
solid lines. The top-right panels show the 2D
distribution of stellar populations within a range of young ages (left panel)
and for ages older than a threshold (right panel), within the structures of interest.
The profiles on the bottom panels correspond to the contribution from stellar populations of differente ages to the
total galaxy light (top), and the age-metallicity relation
(bottom) within each structural component. Although light fractions and metallicities are computed for each discrete age,
a continuous histogram-like profile is shown instead for the sake of clarity. Note there is a significant 
overlap between the light fractions for the outer bar and lens. The vertical solid orange
 and dash-dotted purple lines indicate the age regions used for computing the top averaged maps. 
Stellar populations younger than $\sim$1\,Gyr contribute less than 1\% and they are not shown in the
age-metallicity profile.}
 \label{fig:sfh1291}
\end{figure*}

The information provided by Fig.\,\ref{fig:structuremaps}
is then used to design segmentation maps defining which regions of the MUSE FoV
are dominated (or almost dominated) by each structural component.
The Voronoi binning applied to the data is considered at this stage,
as results within bins and not within spaxels are going to be averaged.
The final segmentation maps for NGC\,1291 and
NGC\,5850 shown in Fig.\,\ref{fig:sfh1291} and Fig.\,\ref{fig:sfh5850}, respectively, are
built by considering a central circular region accounting for PSF
effects and defined with a radius equivalent to 3$\sigma$ of the PSF;
besides, the regions of dominance for each structural component are built
with the selection criteria detailed in Tables\,\ref{tab:structures1291} and
\ref{tab:structures5850}. Some remarks about this selection:

\begin{enumerate}
\item All segments are intentionally chosen to be non-adjacent. In some cases, this is achieved
by imposing tighter constraints
on the light contribution from some structures (see, e.g., the criterion on the bulge light when
selecting the segment corresponding to the inner disc in NGC\,5850).

\item Two bulge regions are considered for NGC\,1291: an \emph{outer bulge},
which is not contaminated by the inner bar (inner bar light contribution $<$0.5\%),
and an \emph{inner bulge}, where the light contribution from the inner bar, while lower
than that from the bulge, is not negligible. 
Given the closeness of NGC\,1291, the central regions are mapped with superb spatial resolution 
and the selection of an inner-bulge segment allows to analyse gradients with respect to the 
outer-bulge and inner-bar regions.

\item The spatial resolution of NGC\,5850 allows to select one bulge segment only. 
As in the outer bulge of NGC\,1291, an almost-zero contribution from the inner bar ($<$0.5\%)
is required in order to avoid any overlapping.

\item The bulge in NGC\,5850 dominates the light over the inner disc only in the very central regions,
where the circular region accounting for PSF effects is located. For that reason, the criterion
for the inner disc when selecting the bulge segment has been relaxed with respect to the other 
structures. 

\item The brightest regions for the inner bar of NGC\,1291 have a similar surface brightness
as the bulge and inner disc in the same bins. This also happens for the outer bar with respect
to the lens.

\item The galaxy disc in NGC\,1291 never dominates the light within the MUSE FoV
and, therefore, no disc segment is defined.
Note however that the underlying disc is to some extent affecting the measurements
for all the structures.

\end{enumerate}

The purpose of avoiding defining adjacent segments is two-folded: first,
since the information of the light contribution from each structure comes from the 2D photometric
decompositions performed on the \sg\ images, it does not necessarily match the light distribution in
the optical range where MUSE works. For this reason we select those regions where the contribution
from each structure is the highest, which will be most likely similar for all bandpasses. Second, 
by making the largest
differentiation between structures we minimise contamination as much as possible,
although we remark that each aperture actually accounts for several overlapping
structures except for, most probably, the region dominated by the
galaxy disc in the case of NGC\,5850. 
 
Once the segmentation maps are created, the results from the \STECKMAP\ analysis
described in Sect.\,\ref{sec:sfh} are averaged
within every structure and compared, as described in 
the following Sect.\,\ref{sec:sfh1291} and \ref{sec:sfh5850}.

\subsection{The star formation history of NGC\,1291 at a glance}\label{sec:sfh1291}
Figure\,\ref{fig:sfh1291} shows the fraction of light contributed by stars of 
different ages, as well as the metallicities, for the different structural components.
The behaviour of the stellar age distribution is similar for all structures: 
between 40\% and 50\% of the light comes from stars that were formed more than 7\,Gyr ago 
(dismissing the central region accounting for PSF
effects, for which the fraction of light from old stars increases up to $\sim$60\%). 
The star formation process smoothly decayed until $\sim$3\,Gyr ago, when it was rapidly quenched such that
stars younger than $\sim$1\,Gyr contribute less than 1\% to the total galaxy light.

While the stellar age distributions for the outermost components, particularly the lens and outer bar,
are exactly the same, a gradient among the innermost structures is noticeable. 
The centre presents the highest contribution from
old stars, which progressively decays for the inner bulge, inner bar, and outer bulge.
These results explain the behaviour observed in the mean luminosity-weighted
ages and metallicities shown in Fig.\,\ref{fig:meanageandmet}, where
the inner bar appears as a slightly older component
than the surrounding structures (inner disc, outer bar, and lens).

Differences among the inner structures (centre, inner and outer bulge, inner bar)
are most apparent in the two top-right subpanels of Fig.\,\ref{fig:sfh1291}, where
the spatial distribution of the total light contributions from stars older than 7\,Gyr and
stars between 1\,Gyr and 2\,Gyr old are shown. This latest age range enhances small differences
($<$10\%) in the contribution from young stars, as the centre and inner bulge appear in blue colours,
the inner bar looks greener, and the outer regions are all coloured in yellow/red.

All structural components show similar metallicity values with a subtle gradient which will
be discussed
in Sect.\,\ref{sec:formation}, where these results will be used to constrain
the dynamical assembly process of NGC\,1291.

\subsection{The star formation history of NGC\,5850 at a glance}\label{sec:sfh5850}
\begin{figure*}
 \vspace{2pt}
 \includegraphics[bb= 10 25 479 440, angle=0.]{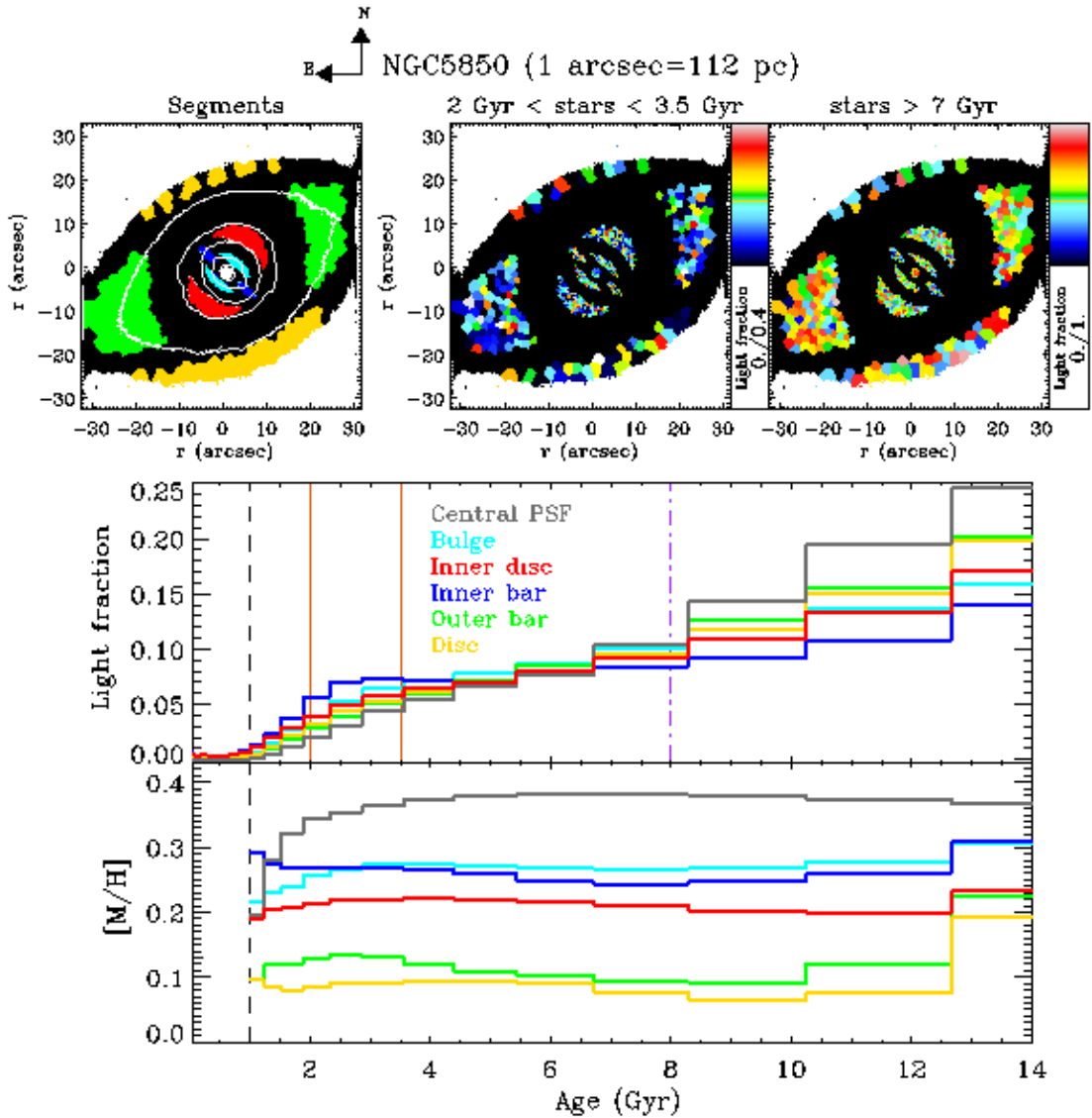}
 \caption{Same as Fig.\,\ref{fig:sfh1291} but for the double-barred galaxy NGC\,5850.}
 \label{fig:sfh5850}
\end{figure*}

Figure\,\ref{fig:sfh5850} shows the corresponding stellar-population analysis for NGC\,5850. 
Significant differences among the inner bar, 
the rest of inner structures (bulge and inner disc),
and the outermost components (outer bar and galaxy disc) are found in this case. 

The most distinct stellar age distribution is that of the inner bar, which shows 
a bump indicating a higher light fraction ($\sim$50\%) coming from stars with ages between 1 and 4.5\,Gyr.
This feature is used to show the different contribution from young stars in the inner bar region
in the top-right subpanels of Fig.\,\ref{fig:sfh5850}; stars older than 7\,Gyr are also
shown, as done for NGC\,1291.

The bulge shows a smoothly declining profile going from light fractions of
$\sim$16\% for very old ages to 6\% for stars of $\sim$3\,Gyr, when a sudden change of slope 
occurs and the profile abruptly decays. Both inner bar and bulge
are more metal-rich than the outermost components (disc and outer bar).

We reiterate here that the inner disc has been
detected from both the photometric decomposition and kinematic analyses in an independent manner. 
Moreover, it is dominant in the bulge region. Indeed both components show a similar
stellar age distribution but with a smaller contribution from
intermediate-age populations (between 4 and 10\,Gyr) in the inner disc, which is also
less metal-rich than the bulge. 

20\% of the outer bar light comes from very old
stars and the light fraction decreases until it reaches $<$1\% due to
stars younger than 1.3\,Gyr. The contribution from stars younger than 1\,Gyr
is indeed negligible for all the structural components, as in NGC\,1291. The stellar age distribution
for the disc is very similar to that of the outer bar, with a subtle rejuvenation noticeable
when comparing ages older and younger than $\sim$6\,Gyr. As seen in Fig.\,\ref{fig:meanageandmet}, the centre
stands out with older and more metal-rich populations. A gradient in metallicities
is observed, from metal-rich to less metal-rich inside outwards.



\subsection{Measurement of the [$\alpha$/Fe] overabundance distribution}\label{sec:alpha}

\begin{figure}
 \includegraphics[bb= 54 -25 320 670, angle=0., width=.26\textwidth]{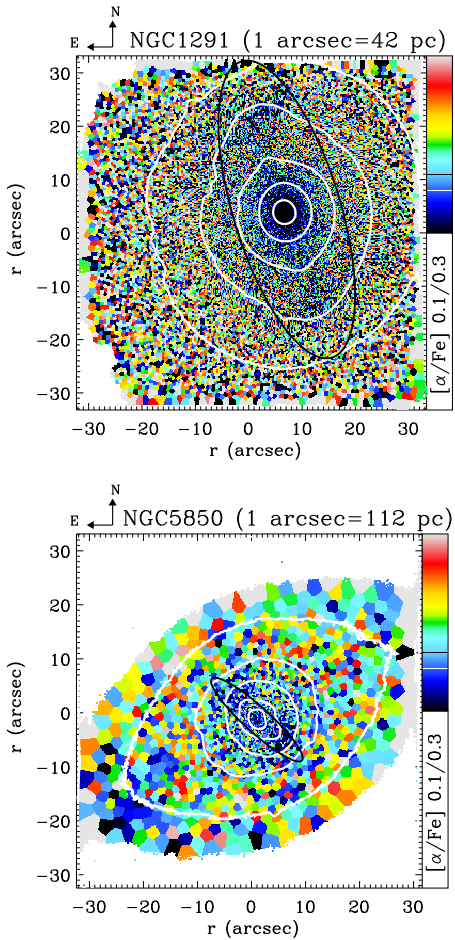}
 \caption{2D spatial distribution of the [$\alpha$/Fe] overabundance for NGC\,1291
(top) and NGC\,5850 (bottom). Typical error bars are of $\sim$0.03\,dex for NGC\,1291 and $\sim$0.11\,dex
for NGC\,5850.
Flux contours are overplotted as solid white 
lines. The inner bar (full length of the Ferrers profile) is outlined in black 
by using the results from the 2D photometric decomposition.}
 \label{fig:alpha}
\end{figure}

In order to retrieve more information about the timescales governing the formation process
for the double-barred galaxies, we derived the 2D spatial distribution of the $[\alpha$/Fe$]$ overabundance.
This was done by using a Bayesian fitting of the measurements of the line-strength
indices with the predictions for the same indices measured
over the solar-scaled and $\alpha$-enhanced MILES-based SSP models of \citet{Vazdekisetal2015}. The line-strength indices used in the fit were
H$\beta_o$ \citep{CervantesandVazdekis2009}, Fe4383 and Fe5015 \citep{Wortheyetal94},
and Fe5270 and Mg$b$ \citep{Bursteinetal1984}. They were selected 
for a best description of the age, metallicity, and $[\alpha$/Fe$]$ within the MUSE spectral range.
This statistical technique has already been successfully used in \citet[][]{MartinNavarroetal2018} 
to retrieve $[\alpha$/Fe$]$ values, as well as single ages and metallicities. 
We refer the reader to that paper for details on how the index-fitting works and the selection
of the indices. 

Figure\,\ref{fig:alpha} shows the $[\alpha$/Fe$]$ 2D distributions for NGC\,1291 and
NGC\,5850.
Inner bars appear with bluer colours than the rest of the FoV, meaning
they have lower $[\alpha$/Fe$]$ values than the outermost structures. 
This result suggests inner bars have undergone a longer formation process than
the inner disc, outer bar, and lens
or galaxy disc, since the timescales
involved in the creation of $\alpha$ elements are shorter than those responsible for Fe formation.


\section{Discussion }\label{sec:discussion}

\subsection{Nature versus nurture of the $\sigma$-hollows}\label{sec:hollows}

\begin{figure*}
 \vspace{2pt}
 \includegraphics[bb= 40 30 620 450, angle=0., width=.75\textwidth]{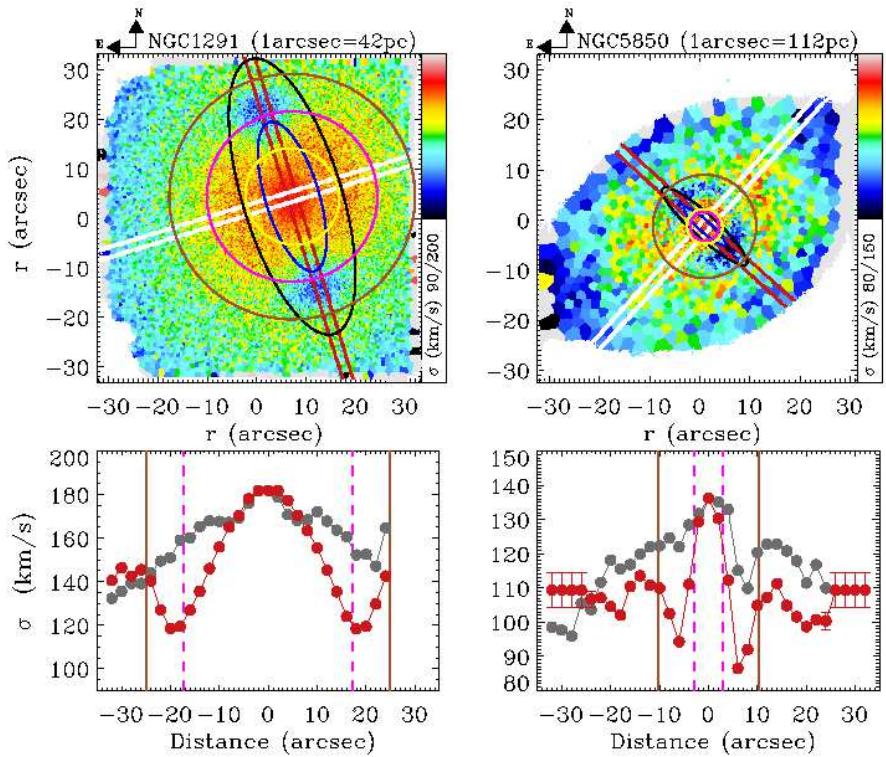}
 \caption{Analysis of the 2D spatial distribution of the stellar velocity dispersion for
NGC\,1291 (left) and NGC\,5850 (right). The top panels show the stellar velocity dispersion maps; the 
projected size and shape of the bulges (as measured with their effective radii; yellow lines) and
inner bars (as measured with both the full length - black lines - and
the effective radii of the corresponding Ferrers profiles - blue lines) are overplotted. Two circles
indicating the radii at which the surface brightness of both bulge and inner bar are the same are
also overplotted (magenta - inner crossing - and brown - outer crossing-).
The $\sigma$-hollows appear within the region where the inner bar dominates over the bulge light. The red and white lines highlight the
regions used as pseudoslits to retrieve the linear $\sigma$ profiles along and perpendicular
to the inner bars, respectively. These profiles (red and grey, respectively) are shown in the bottom panels,
where the radii at which the surface brightnesses profiles of bulges and inner bars cross
are also indicated (dashed magenta line - inner crossing; solid brown line - outer crossing).}
 \label{fig:hollows}
\end{figure*}

The $\sigma$-hollows observed in all double-barred galaxies studied to date 
may provide information about the dynamics of inner bars, but their origin is still unclear. 
\citet{deLorenzoCaceresetal2008} suggested that $\sigma$-hollows are not intrinsic 
features of the velocity dispersion distribution of any isolated structure
(inner bar, disc, or bulge). Within their explanation, inner bars have a rather flat and cold profile
of the velocity dispersion along them, while hot classical bulges show
a decaying distribution from high values, and discs have a smooth axisymmetric distribution
with still higher $\sigma$ values than the inner bar at radii matching with 
the inner bar size. The coexistence of these three structures produces a visual effect
such as local minima due to the low-$\sigma$ values of the inner bar are observed only 
where the inner bar dominates the 
total galaxy light over the hotter bulge and hotter (at those radii) disc, i.e., at the bar ends.
If this hypothesis (which we refer to as \emph{nurture} of the $\sigma$-hollows) was correct,
the appearance of the 
$\sigma$-hollows would depend on the relative contribution of the structures involved
(mainly inner bar and bulge in the previous example) to the total light, as well as 
on their relative level of rotational versus pressure support. Indeed $\sigma$-hollows 
would not be seen in galaxies without a bulge (or other central structure with a different
dispersion value with respect to the inner bar). 

More recently, \citet{Duetal2017a} used numerical simulations to show that 
the presence of any bar, either single or embedded in a double system, 
produces features in the orthogonal velocity dispersion distribution of its
host disc. 
Indeed, the pure $\sigma_z$ of the disc (once isolated from all other structures, including the bar)
shows local minima at the location of the bar ends
and local maxima in the perpendicular direction.
When analysing the total velocity dispersion distribution of a projected 
system composed by a bar and disc, these features resemble the observerd $\sigma$-hollows,
as well as two perpendicular bumps termed as $\sigma$-humps. Simulations
also show that the smaller the bar, the more intense 
both $\sigma$-hollows and humps appear. Why the $\sigma_z$ of the disc is modified by
the presence of a bar and why smaller bars produce more enhanced $\sigma$ features remains unclear.

In this case the presence of a bulge or any other central hotter/cooler structure
than the inner bar is not needed. Indeed, a hot bulge may
mask the detection of these $\sigma$ features, although the $\sigma$-hollows, more prominent than the humps,
usually remain. We reiterate that the disc hosting the bar is responsible for the 
$\sigma$-hollows, which are physically produced in it: its intrinsic velocity dispersion
does decay at those points.

Whether the $\sigma$-hollows are a matter of visual contrast among structures or they have a physical nature
in the galaxy disc is unkown. In an attempt to shed some light on this topic, 
Fig.\,\ref{fig:hollows} shows the 2D velocity dispersion distribution for NGC\,1291 and NGC\,5850
together with the profiles extracted from pseudoslits along the direction of each inner bar, 
and perpendicular to them. The $\sigma$-hollows are clearly observed in the
profiles along the inner bars. The strong inner bar in NGC\,5850 is rather 
narrow in the perpendicular direction; it is therefore complicated to extract conclusions
about possible local maxima for this galaxy, as
predicted by the scenario of \citet{Duetal2017a}. NGC\,1291 does show 
features along the perpendicular direction but at smaller radii, 
and they are better interpreted as
local minima rather than local maxima.

The sizes of the bulges and inner bars for both galaxies have been overplotted in Fig.\,\ref{fig:hollows}.
The corresponding ellipses are drawn by considering the effective radii of the bulges (in yellow), 
the effective radii of the inner bars (in blue), and the full length of the Ferrers profiles for
the inner bars (in black). Moreover, the radii at which the surface brightness from bulge 
and inner bars are the same are indicated with either circles (magenta for the 
innermost crossing, brown for the outermost crossing). This means that the bins inside 
the inner bar and delimited by
the magenta and brown circles are those where the surface brightness of the inner bar is greater
than that of the bulge.

It is noticeable that the $\sigma$-hollows appear at the region where the light coming from the individual
inner bar dominates over the bulge light,
even for the case of NGC\,1291 where the levels of brightness from both structures
are very close, as seen in the inset of Fig.\,\ref{fig:photdec1291}.
This result supports the nurture origin of the $\sigma$-hollows, which are indeed visible where
the inner bar dominates over the presumably hotter bulge and other cooler structures (e.g. inner disc). 

We also emphasise that both galaxies host bulges with relatively-high S\'ersic indices ($n$=2.97 and 2.68
for NGC\,1291 and NGC\,5850, respectively), pointing towards a classical origin for them. 
Classical bulges are expected to be dynamically hot structures. Moreover, in Sect.\,\ref{sec:bulge}
we will introduce the finding of a box/peanut structure at the centre of NGC\,1291, as
found by \citet{MendezAbreuetal2019}; box/peanuts
are thicker than their host bars and therefore they have higher velocity dispersion values
as well. 

Finally,
we warn the reader that, although these evidences support the scenario proposed in 
\citet{deLorenzoCaceresetal2008}, a
physical origin for the $\sigma$-hollows cannot be fully discarded as there is a chance
these regions fortuitously coincide with the alterations of the disc $\sigma_z$ obtained
by \citet{Duetal2017a}.

\subsection{How double-barred galaxies form}\label{sec:formation}

\begin{figure*}
 \vspace{2pt}
 \includegraphics[bb= 54 15 620 600, angle=0., width=.6\textwidth]{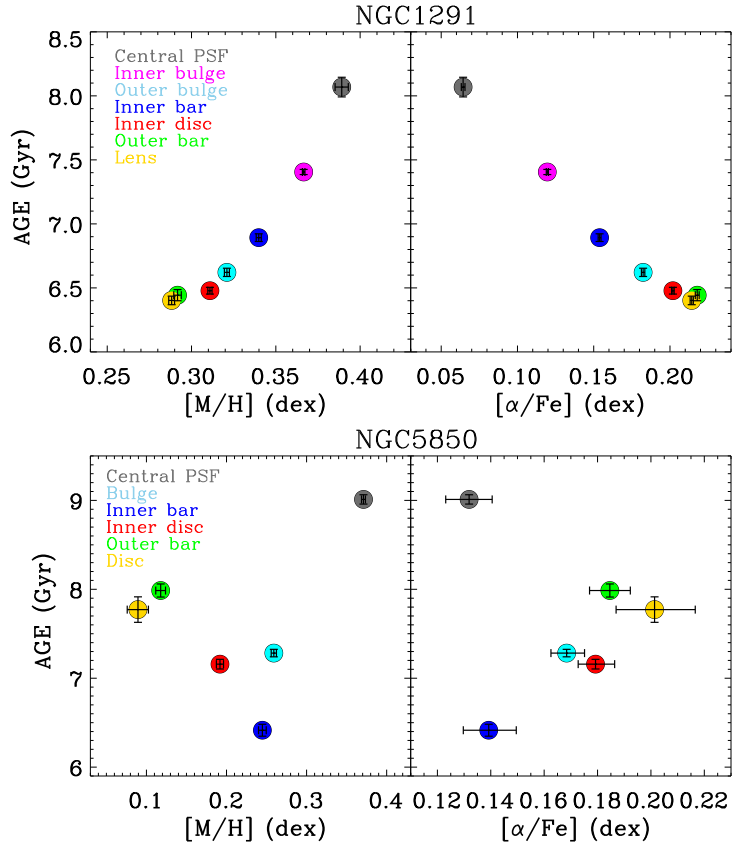}
 \caption{Mean luminosity-weighted age versus metallicity (left panels) and $[\alpha$/Fe$]$ (right panels)
within the spatial regions where each structural component of 
NGC\,1291 (top) and NGC\,5850 (bottom) dominates the light:
inner bulge (magenta, only for NGC\,1291), outer bulge (cyan), 
inner disc (red), inner bar (blue), outer bar (green), and lens (NGC\,5850)/galaxy disc (NGC\,1291; yellow).
Due to possible resolution effects, the central PSF is considered separately (grey). See Sect.\,\ref{sec:segmentationmaps}
for details on the selection of these regions.}
 \label{fig:meansinstructures}
\end{figure*}

Our stellar population results for the individual structural
components in the two double-barred galaxies shed light on the formation path NGC\,1291 and
NGC\,5850 have followed. The first milestone in this discussion is the finding of an inner
disc component for both galaxies, whose sizes match those of the inner bars.
While the inner disc in NGC\,5850 had been already detected by kinematic analysis
\citep[e.g.][]{deLorenzoCaceresetal2013}, the face-on condition of NGC\,1291 prevents
the kinematic identification of the inner disc, that has been detected with the 
2D photometric decompositions instead.

Within the two possible formation scenarios described in Sect.\,\ref{sec:intro}, 
\emph{scenario\,1} invokes a direct formation of the inner bar after gas inflow through
the outer bar. This represents a special way of forming bars, as \emph{normal} bars
(i.e. single bars or large-scale bars within double-barred systems) are formed 
through dynamical instabilities from cold discs 
\citep[e.g. ][]{Combesetal90,DebattistaandSellwood2000,Athanassoula2003}.
Inner and outer bars may therefore be considered intrinsically different
within this scenario. It is worth noting that inner bars do not usually present
gaseous counterparts \citep[][but see the case of NGC\,2273 in \citealt{PetitpasandWilson2002}]
{PetitpasandWilson2004}. 

On the other hand, in \emph{scenario\,2} both inner and outer (or single) bars 
follow the same formation process, i.e., both are born from instabilities 
in dynamically cold discs. While large-scale bars are formed in galaxy discs,
inner bars are embedded in inner discs. Inner discs are a rather frequent
structural component, appearing in $\sim$29\% of ellipticals and S0s
\citep[][estimate from kinematic analyses]{Krajnovicetal2008}. 
This scenario therefore considers that all
bars share the same nature and they may therefore undergo similar evolutionary
processes, which may be different only when differences in size play a role.

The discovery of inner discs in our two galaxies supports this second scenario, further
backed by the fact that the inner bar in NGC\,1291 appears older than the outer bar.
This possibility is not accounted for in \emph{scenario\,1}, where the outer bar needs to 
be formed prior to the inner bar, while \emph{scenario 2} does not impose any 
sequential requirement.
The fact that the inner discs may be very faint, as in the case of NGC\,1291, 
would explain why they have not been detected in all double-barred galaxies 
observed. We remind here that most of the studies available for double-barred
galaxies are photometrically based; however, the spectroscopic analyses of 
\citet{deLorenzoCaceresetal2012}
and \citet{deLorenzoCaceresetal2013} do find kinematic signatures of inner discs
in 4 out of 5 double-barred galaxies under study. The remaining galaxy hosts
a very small inner bar and resolution effects may have prevented the detection
of the inner disc in this case.

The measurements shown in Sect.\,\ref{sec:sfh}
indicate that NGC\,1291 and NGC\,5850 have undergone
formation processes with different pathways and therefore
they are discussed separately as follows.

\subsubsection{NGC\,1291: a rather contemporary assembly of all structures at an early epoch}

NGC\,1291 shows similar SFH for all the structures in place within the MUSE
FoV; from inside outwards: centre, bulge, inner bar, inner disc, outer bar, and lens (Fig.\,\ref{fig:sfh1291}).
Figure\,\ref{fig:meansinstructures} shows the mean luminosity-weighted ages, 
metallicities, and $[\alpha$/Fe$]$ linearly averaged within the regions of predominance of
each individual structure. It is important
to remark here that, although each component is dominant inside its own segment, its contribution
cannot be fully isolated with respect to the rest of overlapping structures such as, at least, the galaxy
disc.

The errors associated to the mean values shown in Fig.\,\ref{fig:meansinstructures} have been computed
as the quadratic sum of the uncertainties in the age and metallicity measurements (typically
of 0.9\,Gyr and 0.06\,dex, respectively) and the standard
deviation of the values which are averaged for each structure, corrected by the number of 
elements involved in the average. 

The inside-out gradient already pointed out in Sect.\,\ref{sec:sfh1291} is clearly 
seen in Fig.\,\ref{fig:meansinstructures} not only in age but also in metallicity and
$[\alpha$/Fe$]$. The centre (represented by the central PSF and the inner bulge)
is the oldest, more metal-rich, and less $\alpha$-enhanced part of the galaxy.
The remaining structures (outer bulge, inner bar, inner disc, outer bar, and lens) show old and
particularly similar ages: values expand
from 6.4 to 6.9\,Gyr, and also the metallicity
and $[\alpha$/Fe$]$ cover short ranges. 
This result points towards a rapid consumption of
most of the gas available for triggering significant star formation, so most of the stars, 
now distributed
along the whole
galaxy, were formed at an early stage.

So far we have discussed when the stars in NGC\,1291 were formed.
In the following we will discuss how the morphological assembly of this galaxy took place.
We emphasise here that the assembly of structures may or may not involve star formation
and therefore the analysis of SFH in these regions does not necessarily imply 
the formation history of the structures. However, the combination of SFH and the 
kinematic and structural analysis presented here is a powerful approach to constrain
the assembly history of NGC\,1291.

Within \emph{scenario\,2}, the bar structures in NGC\,1291 were born from 
dynamical instabilities of the outer and inner disc. The inner disc was secularly formed after
gas inflow along the outer bar. This process does involve star formation: the fact that the inner
disc has the same age as the outer galaxy regions (outer bar and lens) indicate that the whole process
of outer bar formation, gas inflow, and inner disc formation was fast. This explains that no rejuvenation of the
inner disc with respect to the outer bar is noticeable but the inner disc actually is
slightly more metal-rich and $\alpha$-enhanced, as expected.

Once the inner disc is formed, the dynamical assembly of the inner bar happened at a later
stage from the same stellar content. The fact that the inner bar is older than the inner disc
is explained by invoking some level
of star formation happening in the inner disc (as well as in the outer regions) 
after the inner bar was already assembled
\citep[as it has actually been seen in simulations and observations; ][]{Athanassoula92,Emsellemetal2015,RomeoandFathi2016}. 
This is not surprising: the shear exerted by
the inner bar prevents star formation inside it, while the gas still present in the galaxy may
continue forming stars outside.

The latest result also agrees with the fact that the outer bulge region, which is purposedly taken where
the contribution from the inner bar is negligible, is also younger than the inner bar. 
Only structures
inside the inner bar remain the oldest, as the shear introduced
by the inner bar prevents star formation and subsequent rejuvenation.
We can also pose constraints on the moment of assembly of the inner bar: it had to take place before the slight
rejuvenation of the inner disc and outer structures happened, i.e., $>$6.5\,Gyr ago.

Important conclusions are inferred from this assembly history,
as it implies this inner bar is a long-lived structure. This conclusion is further reinforced
by the presence of a box/peanut associated to the inner bar of NGC\,1291, as it will be explained in Sect.\,\ref{sec:bulge}
\citep[see also][]{MendezAbreuetal2019}.
The transient versus long-living nature of inner bars 
has been a matter of debate for long time since many numerical simulations, mostly those
corresponding to \emph{scenario\,1}, generate fragile inner bars that live for very short periods
of time (dissolving in few hundreds of Myr).

Regarding the old and metal-rich populations at the centre of NGC\,1291 (central PSF, inner bulge),
they need to be explained through
an early formation of these very central structures either in a very intense way with a high star formation
rate which rapidly enriched
the medium, either from an already enriched reservoir of gas everywhere in the galaxy.
This second option implies that an external supply of less metal-rich gas 
has to be invoked to explain the fuel for forming the rest of the components.
We remind that 
NGC\,1291 belongs to a group \citep{HuchraandGeller1982}, but gas inflow from the galaxy
halo can also explains these results.

\subsubsection{NGC\,5850: an inner bar assembled more than 4.5\,Gyr ago}
The inner and outer structures within NGC\,5850 are clearly distinguishable in their stellar populations. 
In this case, the inner bar is younger, more metal-rich,
and less $\alpha$-enhanced
than the outer bar. This result matches with predictions from \emph{scenario\,1}, as already pointed
out in \citet{deLorenzoCaceresetal2013}. However, it also fits within \emph{scenario\,2} and 
this second possibility is reinforced by the presence of the inner disc.

The results for NGC\,5850 indicate its inner bar
has a major contribution from stars between 1 and 4.5\,Gyr old. Within \emph{scenario\,2},
this inner bar was dynamically assembled from the inner disc, but this is older, less metallic, 
and more $\alpha$-enhanced. These results can be easily reconciled if star formation
occurred along the inner bar after it was assembled from the inner disc; such a star-forming
period took place between 4.5 and 1\,Gyr ago.
This is not surprising for an inner bar as strong as that of NGC\,5850
\citep[the inner
bar in NGC\,5850 shows $\epsilon\sim$0.8 and 
the ellipticity is a proxy for bar strength, as indicated by e.g.][]{Laurikainenetal2002}. 
When such a strong bar is formed, 
gas quickly reacts to its potential
and star formation is triggered inside the barred structure 
\citep[e.g. ][]{ElmegreenandElmegreen85, FriedliandBenz95, MartinandFriedli1997}. Such behaviour 
is not expected in NGC\,1291, which hosts a weaker inner bar ($\epsilon\sim$0.6)
and therefore its shear produces the opposite effect of preventing star formation.

Since the rejuvenation of the inner bar with respect to the inner disc had to happen
after the inner bar was dynamically assembled, the evidences shown in Fig.\,\ref{fig:sfh5850}
indicate that it was fully formed $>$4.5\,Gyr ago. As in NGC\,1291, this result
supports a long-living nature for the inner bar of NGC\,5850. We note here that 
\citet{deLorenzoCaceresetal2012, deLorenzoCaceresetal2013} also suggested inner bars
are long-lived structures when studying four additional double-barred galaxies.
To the best of our knowledge, no observational evidences of short-lived bars have been found so far.

As expected, the outer regions (outer bar and disc) of NGC\,5850 are old, with a slight difference 
in age so the outer bar is slightly older than the disc. As discussed for NGC\,1291, this evidence may
indicate that star formation was inhibited within the bar region once it is formed,
while it continues happening within the disc.
NGC\,5850 indeed hosts a galaxy disc with spiral arms where star formation is noticeable.
We emphasise, however, that the MUSE FoV probes the NGC\,5850 innermost disc regions, 
at the same galactic longitudes
where the bar exists. 

The central PSF of NGC\,5850, a galaxy in a pair \citep{Madoreetal2004},
shows the same very old, more metal-rich, and less $\alpha$-enhanced
behaviour as NGC\,1291. The bulge in this case corresponds to a region where the contribution from 
the inner bar is negligible (like the \emph{outer bulge} in NGC\,1291) and 
it shows stellar-population properties closer to those of the inner disc. 
Such very old and metal-rich central regions are found for many TIMER galaxies and will be discussed
in a forthcoming paper (S\'anchez-Bl\'azquez et al. in preparation).

\subsection{Secular evolution driven by inner bars}\label{sec:bulge}
Bars are considered the main drivers of secular evolution as, thanks to their
non-axisymmetric potential, they can promote stellar radial migration
outwards and gas inflow which eventually may trigger star formation.
Inner bars are also considered capable of such an effect. They have even been proposed as 
a mechanism to bring gas closer to the sphere of influence
of the supermassive black holes \citep{Shlosmanetal89, Shlosmanetal90}; gas transported by a single bar cannot reach
distances less than $\sim$100\,pc from the centre. This idea of secular evolution promoted by inner bars is further
supported by \emph{scenario 2}, which implies that all bars have the same nature.
However, no evidence of any rejuvenation due to gas inflow through inner bars is
found in these galaxies; in \citet{deLorenzoCaceresetal2012, deLorenzoCaceresetal2013}
similar results were found, as well as signatures of gas inflow from the outer
to the inner bars. Gas may indeed be inflowing but not be forming stars in a significant
way. The analysis of the ionised gas in the TIMER double-barred galaxies
will be presented in a forthcoming paper.

Numerical simulations have shown that buckling instabilities in bars
are responsible for the formation of the so-called box/peanut structures at galaxy centres
\citep[e.g.][and references therein]{MartinezValpuestaetal2006}. This secular process promoted by bars
is based on the redistribution of existing stars and does not
imply new star star formation, i.e., different stellar population content for the box/peanut
with respect to the bar is not expected. 
Whether inner bars may drive box/peanut formation as well has not been assessed in the literature so far.
In a companion paper 
\citep{MendezAbreuetal2019} we demonstrate 
that the inner bar of NGC\,1291 does show the expected kinematic signatures of a box/peanut 
seen face-on in its h$_4$ distribution \citep[see ][for numerical and observational evidence, 
respectively]{Debattistaetal2005,MendezAbreuetal2008II}. 

Estimates from numerical simulations indicate that box/peanuts are typically born 
$\sim$1\,Gyr after the formation of the bar. Whether this timescale holds for small-scale bars
has not been studied yet and is debatable. If so, this would be a further confirmation of the long-lived
nature of inner bars. Finally, the presence of a box/peanut structure, so far associated
to large-scale bars alone, reinforces the idea of inner and outer bars having the same nature.
This conclusion once again matches with the hypothesis of \emph{scenario\,2}.
We refer the reader to \citet{MendezAbreuetal2019} for more discussion on these
topics.

The nature of the bulge in NGC\,5850 is not so clear as in NGC\,1291. The bulge component
has a S\'ersic index of 2.68 as derived from the photometric decompositions presented in 
Sect.\,\ref{sec:photdec}. This would correspond to a classical bulge under the criterion
of \citet{FisherandDrory2008}, but the use of this photometric diagnostic
for retrieving the bulge nature 
has  recently been challenged by several authors \citep[e.g.][]{Costantinetal2017,MendezAbreuetal2018}.
In any case and as discussed before, we do not find any evidence of recent star formation bursts
thanks to gas inflow along the inner bar. What we do find is that the central region (central PSF
in the plots) does show lower $[\alpha$/Fe$]$ values, 
pointing towards a more extended star formation process; intriguinly, it also has the oldest 
stellar populations.


\section{Summary and conclusions}\label{sec:conclusions}
TIMER is a project devoted to the detailed study (gas and stellar content) of 
24 barred galaxies hosting presumably secularly-formed inner structures that have been
observed with the MUSE IFU.
In this paper, we present the analysis of the two double-barred galaxies included in the TIMER sample:
NGC\,1291 and NGC\,5850. Given the structural complexity of double-barred systems, we have first performed
2D multi-component photometric decompositions of the \sg\ images for these galaxies, in order 
to disentangle the mix of structures lying at the central regions covered by the MUSE FoV. A careful
analysis of the stellar properties (kinematics and stellar populations) of these structures is
carried out throughout the paper. The main observational results obtained are:
\begin{itemize}
\item Both galaxies host a stellar inner disc matching in size with the corresponding inner bar. While
the inner disc in NGC\,5850 had already been noticed through kinematical analysis, the one in NGC\,1291
(a face-on galaxy) has been unveiled by our 2D photometric decomposition.

\item The presence of $\sigma$-hollows, two local decreases of the stellar velocity dispersion values
at the ends of inner bars \citep{deLorenzoCaceresetal2008}, 
is confirmed for the two double-barred galaxies analysed here. 

\item NGC\,5850 shows a ring-like feature with high h$_4$ values as expected for inner bars 
\citep{Duetal2016}. The h$_4$ distribution for NGC\,1291 is more complicated and it is
the subject of a companion paper \citep{MendezAbreuetal2019}.

\item Inner bars show clearly distinct stellar populations than the surrounding regions. In NGC\,5850,
the inner bar is younger and more metal-rich, confirming the results by \citet{deLorenzoCaceresetal2013}.
On the contrary, NGC\,1291 hosts an older and more metal-rich inner bar than the outer structures
(inner disc, outer bar, lens, and even the outermost part of the bulge, outside the inner bar region).

\item The stellar age distribution and age-metallicity relation for the dominant structural components
of NGC\,1291 within the MUSE FoV, namely bulge, inner bar, inner disc, outer bar, and lens,
are rather similar. 

\item The star formation history for the inner bar of NGC\,5850 shows a significant bump
between 1 and 4.5\,Gyr. All other structures (bulge, inner disc, outer bar, and disc)
have smoothly declining profiles.

\item Both inner bars show lower [$\alpha$/Fe] values than the outer regions.

\end{itemize}

The above observational pieces of evidence have been discussed within the frameworks of the two main
formation scenarios proposed for inner bars: \emph{scenario\,1}, in which the inner bar is directly assembled
through a gas-rich process that implies in-situ star formation in the inner bar
\citep[e.g.][]{FriedliandMartinet93}. According to simulations,
this scenario usually forms fragile inner bars; and \emph{scenario\,2},
in which the origin of inner bars is exactly the same as that for outer bars, i.e. formation
through dynamical instabilities in cold discs, but at smaller spatial scales 
\citep[e.g.][]{DebattistaandShen2007}. This scenario has been successful in simulating
long-lived inner bars.
The main conclusions derived from the analysis presented here regarding the formation
and evolution of inner bars  are:

\begin{itemize}

\item Given the presence of inner discs in the majority of double-barred galaxies studied in detail
\citep[see also the works by ][]{deLorenzoCaceresetal2012, deLorenzoCaceresetal2013};
and the results obtained for NGC\,1291 (inner bar appears slightly older than outer bar),
we suggest \emph{scenario\,2} is better suited to explain 
how double-barred galaxies form. 

\item The fact that  all structures in NGC\,1291 present similar old ages and metallicities suggest
this galaxy formed all stars and assembled all components
(as the inner disc is most probable the result of a star-forming process) within a short timescale.

\item The rejuvenation of the inner disc in NGC\,1291 with respect to the inner bar
may be due to a slightly longer 
star formation process in the inner disc after the inner bar was formed. The star formation
in the \emph{weak} ($\epsilon\sim$0.6) inner bar would thus have been quenched when it was assembled, as predicted by 
\citet{Athanassoula92} and \citet{Emsellemetal2015} 
for large-scale bars. This result constrains the 
moment of assembly of the inner bar to an early-epoch, so its mean age 
is older than 6.5\,Gyr. 

\item NGC\,5850 hosts a strong inner bar ($\epsilon\sim$0.8) for which star-formation triggering
right after inner-bar assembly is expected, 
as predicted by e.g. \citet{FriedliandBenz95}. This rejuvenation process, further supported 
by the younger and more metal-rich stellar
content of this inner bar, took place between 1 and 4.5\,Gyr ago. 
Therefore, the inner bar was assembled before it could be rejuvenated with respect to the inner disc,
i.e., $\sim$4.5\,Gyr ago.

\item The assembly epochs for the inner bars derived in this work and the fact that the inner bar
NGC\,1291 has developed a box/peanut structure \citep{MendezAbreuetal2019}
backs the hypothesis that inner bars are 
long-lived structures, as suggested by several numerical works \citep[e.g.][]{Wozniak2015}.

\item We find no significantly younger stellar populations for the bulges with respect to the inner bars,
nor signatures of more recent star formation after the inner disc formation that could be due
to secular evolution promoted by inner bars. 

\end{itemize}

In summary, by combining a structural analysis with a star formation history analysis and thanks
to the superb properties of the TIMER MUSE data, we have 
for the first time clocked the dynamical assembly history of double-barred galaxies, 
which supports the facts that (i) inner bars are dynamically formed from discs in 
an analogous way as outer bars; and (ii) 
inner bars are long-lived structures.

\section*{Acknowledgments}
We are grateful to Peter Erwin for providing his HST data for NGC\,4984 for inspection,
as well as to the anonymous referee for many valuable comments which helped improving
this manuscript.
AdLC and JMA thank the European Southern Observatory for the warm hospitality 
during a science visit when part of this work was done.
This work is based on observations collected at the European Organisation for Astronomical Research 
in the Southern Hemisphere under ESO program 097.B-0640(A).
AdLC acknowledges support from grant AYA2016-77237-C3-1-P from the Spanish 
Ministry of Economy and Competitiveness (MINECO).
PSB acknowledges support from grant AYA2016-77237-C3-2-P from the Spanish Ministry 
of Economy and Competitiveness (MINECO). JMA acknowledges support from 
the Spanish  Ministry of Economy
and Competitiveness (MINECO) by grants AYA2013-43188-P and AYA2017-83204-P.
GvdV acknowledges funding from the European Research Council (ERC) under the European 
Union's Horizon 2020 research and innovation programme under grant agreement No 724857 
(Consolidator Grant ArcheoDyn).

\bibliographystyle{mn2e}
\bibliography{reference}



\bsp

\label{lastpage}

\end{document}